\documentclass[aps, pre, superscriptaddress, twocolumn, showpacs, amsmath, longbibliography, floatfix]{revtex4-1}

\usepackage{dcolumn}
\usepackage{bm} 
\usepackage{graphicx} 
\usepackage{color}
\usepackage{subfigure}
\usepackage{hyperref}
\usepackage{latexsym}
\usepackage{amsthm}
\usepackage{amssymb}
\usepackage{verbatim}
\usepackage{wasysym}

\DeclareGraphicsExtensions{.jpg, .pdf, .mps, .png, .eps, .ps, .EPS, .gif}
\begin{document}
\newcommand{\red}[1]{#1}

\title{Bond percolation on simple cubic lattices with extended neighborhoods}

\author{Zhipeng Xun}
\email{zpxun@cumt.edu.cn}
\affiliation{School of Materials and Physics, China University of Mining and Technology, Xuzhou 221116, China}
\author{Robert M. Ziff}
\email{rziff@umich.edu}
\affiliation{Department of Chemical Engineering and Center for the Study of Complex Systems, University of Michigan, 
Ann Arbor, Michigan 48109-2800, USA}

\date{\today}

\begin{abstract} 
We study bond percolation on the simple cubic (sc) lattice \red{with various combinations of first, second, third, and fourth nearest-neighbors} by Monte Carlo simulation. Using a single-cluster growth algorithm, we find precise values of the bond thresholds. Correlations between percolation thresholds and lattice properties are discussed, and our results show that the percolation thresholds of these and other three-dimensional lattices decrease monotonically with the coordination number $z$ quite accurately according to a power law $p_{c} \sim z^{-a}$, with exponent $a = 1.111$.  However, for large $z$, the threshold must approach the Bethe lattice result $p_c = 1/(z-1)$.  Fitting our data and data for lattices with additional nearest neighbors, we find $p_c(z-1)=1+1.224 z^{-1/2}$.

\end{abstract}


\maketitle
\section{Introduction}

Percolation is a fundamental model in statistical physics \cite{BroadbentHammersley1957,StaufferAharony1994}. It is used to describe a variety of natural processes, such as liquids moving in porous media \cite{BolandtabaSkauge2011,MourzenkoThovertAdler2011}, forest fire problems \cite{Henley1993,GuisoniLoscarAlbano2011} and epidemics \cite{MooreNewman2000}. It is also a model \red{for} phase-transition phenomena. In percolation systems, sites or bonds on a lattice are occupied with probability $p$, and the value of $p$ at which an infinite cluster (in an infinite system) first appears is known as the percolation threshold $p_{c}$.

Many kinds of lattices, graphs, and networks have been investigated to find the percolation thresholds and the corresponding critical exponents. In two dimensions, exact values of percolation thresholds are known for several classes of lattices \cite{SykesEssam1964,SudingZiff1999,Wierman84,Scullard2006,Ziff2006,Wu06,ZiffScullard2006,ZiffSapoval86,DamavandiZiff15},
but there are still many more lattices where thresholds cannot be found analytically, and in higher dimensions there are no exact solutions at all.  Consequently, \red{a main focus} of investigation at present is still based on approximation schemes or numerical simulations.

Numerous algorithms and techniques have been developed to find threshold numerically \cite{,SykesEssam1964-2,HoshenKopelman1976,ReynoldsStanleyKlein80,StaufferAharony1994,LorenzZiff1998,ZiffSapoval86,NewmanZiff2000,DerridaDeSeze82,Jacobsen14,FumikoShoichiMotoo1989,WangZhouZhangGaroniDeng2013,XuWangLvDeng2014,MertensZiff16,SampaioFilhoEtAl18}.
Many related problems in percolation have also received attention recently \cite{KryvenZiffBianconi19,RamirezCentresRamirezPastor19,GschwendHerrmann19,Koza19,MertensMoore2018,MertensMoore18s,ScullardJacobsen2019,TarasevichEserkepov19,MertensMoore19,AppertRollandHilhorst19}\red{---it remains a very active field}.

The study of three-dimensional lattices (the most common ones being the simple cubic (sc), the face-centered cubic (fcc), the body-centered cubic (bcc), and diamond lattices) is particularly important, due to their relevance for many natural processes. Much work in finding thresholds and critical exponents has been done in three dimensions \cite{Marck1997,LorenzZiff1998,WangZhouZhangGaroniDeng2013,XuWangLvDeng2014,Marck1998,vanderMarck98,DammerHinrichsen2004,KurzawskiMalarz2012,Malarz2015,MertensMoore2018}, and the values of percolation thresholds have been more and more accurate. Lorenz and Ziff \cite{LorenzZiff1998} performed extensive Monte Carlo simulations to study bond percolation on three-dimensional lattices ($p_{c}(\rm sc)=0.2488126(5)$, $p_{c}(\rm fcc)=0.1201635(10)$, and $p_{c}(\rm bcc)=0.1802875(10)$) using an epidemic cluster-growth approach. By examining wrapping probabilities, Wang et al.\ \cite{WangZhouZhangGaroniDeng2013} and Xu et al.\ \cite{XuWangLvDeng2014} also carried out extensive numerical simulation studies on these models and found $p_{c}(\rm sc)=0.24881185(10)$, $p_{c}(\rm fcc)=0.12016377(15)$, $p_{c}(\rm bcc)=0.18028762(20)$ for bond percolation, and $p_{c}(\rm sc)=0.31160768(15)$, $p_{c}(\rm fcc)=0.19923517(20)$, $p_{c}(\rm bcc)=0.2459615(2)$ for site percolation, as well as investigating critical exponents.  In general, pure Monte Carlo results are practically limited to about eight significant digits of accuracy, due to statistical error and limitations of computers: at least $10^{16}$ random numbers must be generated to achieve that level of accuracy, and would require $\approx 10^4$ days of computation on a single node.

\begin{figure}[htbp] 
\centering
\includegraphics[width=3.5in]{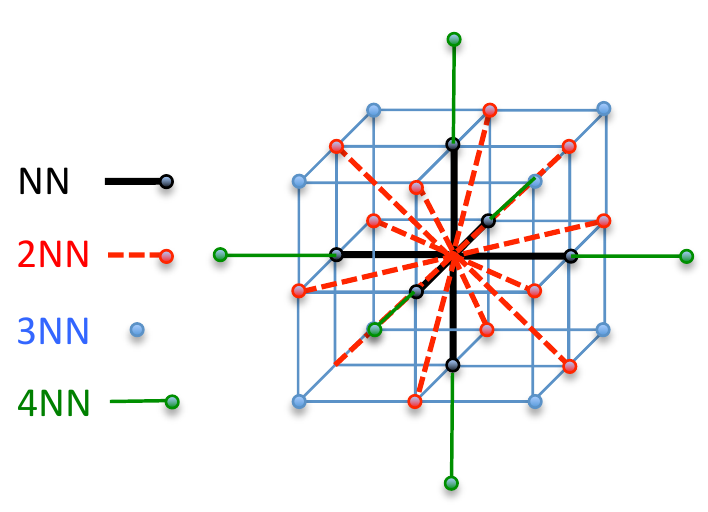} 
\caption{ The neighborhoods considered here: nearest-neighbors (NN) \red{(black with heavy bond, 6 vertices)}; 2NN \red{(red with dashed bond, 12 vertices)}; 3NN \red{(blue with no links to the origin, 8 vertices)}; and 4NN \red{(green with thin bond, 6 vertices)}.}
\label{FigXunZiff}
\end{figure}

The problem of studying percolation on lattices with extended neighborhoods has received a great deal of attention in the last decades \cite{PetrovStoynevBabalievskii91,MajewskiMalarz2007,KurzawskiMalarz2012,Malarz2015,KotwicaGronekMalarz19}, with much work stimulated by the 2005 paper of Malarz and Galam \cite{MalarzGalam05}.  With extended neighbors, the coordination
number $z$ can be varied over a wide range, so many types of systems can be studied, and also there are
applications where these results are useful \cite{Bianconi2013}.  Site percolation on lattices with extended neighborhoods corresponds to problems of adsorption of extended shapes on a lattice, such as $k\times k$ squares on a square lattice \cite{KozaKondratSuszcaynski14,KozaPola2016}.  Bond percolation relates to long-range links similar to small-world networks \cite{Kleinberg00} and models of long-range percolation \cite{SanderWarrenSokolov03}. In two dimensions, having lattices with complex neighborhoods models non-planar systems.

For three-dimensional systems, some work has been done for the sc lattice with extended neighborhoods \cite{KurzawskiMalarz2012,Malarz2015}, although to relatively low precision and for site percolation only. Precise percolation thresholds are  needed in order to study the critical behavior, including critical exponents, critical crossing probabilities, critical and excess cluster numbers, etc. Therefore, in this paper,  we study bond percolation for several sc lattices with extended neighborhoods, including combinations of nearest-neighbors (NN), second nearest-neighbors (2NN), third nearest-neighbors (3NN), and fourth nearest-neighbors (4NN), as shown in Fig.\ \ref{FigXunZiff}. We use
an effective single-cluster growth method similar to that of Lorenz and Ziff  \cite{LorenzZiff1998} and what we have recently used to study percolation problems in four dimensions \cite{XunZiff2019}.  Thresholds for these \red{systems} were never studied for bond percolation, as far as we know, and thus we find all new values.  We find results to a precision of five or six significant digits.

With regard to the sc lattice with extended neighborhoods, crossing bonds exist in \red{these kinds of structures}. This bond percolation model with crossing bonds lives in an extended space of connectivities \cite{FengDengBlote2008}. 
Here we show that the single-cluster growth method we used in this paper can be efficiently applied to \red{these kinds of lattices}.

Another goal of this paper is to explore the relation between percolation threshold and coordination number. The value of percolation thresholds depends on kind of percolation (site or bond), lattice topology and assumed neighborhoods, etc. 
The study of how thresholds depend upon lattice structure, especially the coordination number $z$, has also had a long history \cite{ScherZallen70,GalamMauger96,vanderMarck97,Wierman02,WiermanNaor05,Wierman2017}. Having thresholds of more lattices is useful for extending those correlations.

In the following sections, we present the underlying theory, and discuss the simulation process. Then we present and briefly discuss the results that we obtained from our simulations.

\section{Theory}\label{sec:model}
A quantity of central interest in percolation is the cluster size distribution $n_{s}(p)$, which is defined as the number of clusters (per site) containing $s$ occupied sites, as a function of the occupation probability $p$. At the percolation threshold $p_{c}$, $n_{s}$ is expected to behave as
\begin{equation}
n_{s} \sim A_0 s^{-\tau} (1+B_0 s^{-\Omega}+\dots),
\end{equation}
where $\tau$ is the Fisher exponent, and $\Omega$ is the exponent for the leading correction to scaling. Both $\tau$ and $\Omega$ are expected to be universal---the same for all lattices of a given dimensionality. In three dimensions, relatively accurate results for $\tau$ exist: $2.18906(8)$ \cite{BallesterosFernandezMartin-MayorSudupeParisiRuiz-Lorenzo1997} and $2.18909(5)$ \cite{XuWangLvDeng2014}.  For $\Omega$, the value is not known to comparable accuracy: $0.64(2)$ \cite{LorenzZiff1998}, $0.65(2)$ \cite{GimelNicolaiDurand2000}, $0.60(8)$ \cite{Tiggemann2001}, $0.64(5)$ \cite{BallesterosFernandezMartin-MayorSudupeParisiRuiz-Lorenzo1999}, and a recent higher value, $\Omega = 0.77(3)$ \cite{MertensMoore2018}.  The $A_0$ and $B_0$ are constants that depend upon the system and are non-universal.

\red{Note that even though we are considering bond percolation, we characterize the size of the cluster by the number of sites it contains.  This is in fact a common way to do it, and convenient for the growth method to generate clusters that we employ here, where we do not determine the states of internal bonds.  This is also natural in many theoretical approaches such as the Temperley-Lieb calculation for percolation \cite{TemperleyLieb71}.  In any case, the number of occupied bonds of a cluster is proportional to the number of occupied sites for large clusters, so either choice will yield the same scaling.}

The probability a site (vertex) belongs to a cluster with size greater than or equal to $s$ will then be
\begin{equation}
P_{\geq s} = \sum_{s'=s}^\infty s' n_{s'} \sim A_1s^{2-\tau} (1+B_1s^{-\Omega}+\dots),
\label{ps}
\end{equation}
where $A_1 = A_0/(\tau-2)$ and $B_1 = (\tau-2)B_0/(\tau+\Omega-2)$.
When the probability $p$ is away from $p_{c}$, a scaling function needs to be included. Then the behavior for large $s$ (ignoring corrects to scaling here) can be represented as
\begin{equation}
P_{\geq s} \sim A_2 s^{2- \tau} f(B_2(p-p_{c})s^{\sigma}),
\label{ps2}
\end{equation}
Here $\sigma$ is another universal exponent, which is estimated to be $0.4522(8)$ \cite{BallesterosFernandezMartin-MayorSudupeParisiRuiz-Lorenzo1997}, $0.45237(8)$ \cite{XuWangLvDeng2014}, and $0.4419$ \cite{Gracey2015}.

The scaling function $f(x)$ can be expanded as a Taylor series,
\begin{equation}
f(B_2(p-p_{c})s^{\sigma}) \sim 1+C_2(p-p_{c})s^{\sigma}+ \cdot\cdot\cdot.
\label{scaling}
\end{equation}
where $C_2 = B_2 f'(0)$.  We assume $f(0)=1$, so that $A_2$ = $A_1$. Combining Eqs.\ (\ref{ps2}) and (\ref{scaling}) leads to
\begin{equation}
s^{\tau-2}P_{\geq s} \sim A_2+D_2(p-p_{c})s^{\sigma}.
\label{vssigma}
\end{equation}
where $D_2=A_2 C_2$.

The theory mentioned  above  provides us two methods to determine $p_{c}$. The first way, we can plot $s^{\tau-2}P_{\geq s}$ vs $s^{\sigma}$. Equation (\ref{vssigma}) predicts that $s^{\tau-2}P_{\geq s}$ will convergence to a constant value at $p_{c}$ for large $s$, while it deviates from a constant value when $p$ is away from $p_{c}$. The second way, we can plot $s^{\tau-2}P_{\geq s}$ vs $s^{-\Omega}$. It can be seen from Eq.\ (\ref{ps}) that there will be a linear relationship between $s^{\tau-2}P_{\geq s}$ and $s^{-\Omega}$ for large $s$, if we choose the correct value of $\Omega$, while for $p \ne p_c$, \red{where Eq.\ (\ref{ps}) does not apply,} the behavior  will be nonlinear.

We also consider a third method to study $p_c$ and $\tau$.  It follows from Eq.\ (\ref{ps}) that, at $p_c$,
\begin{equation}
\begin{aligned}
\frac{\ln P_{\geq 2s} - \ln P_{\geq s}}{\ln 2} &\sim \frac{(2 - \tau)(\ln 2s - \ln s)}{\ln 2} - \frac{B_1 s^{-\Omega}(2^{-\Omega}-1)}{\ln 2} \\
&\sim (2 - \tau) + \red{B_3} s^{-\Omega},
\end{aligned}
\label{localslope}
\end{equation}
where $(\ln P_{\geq 2s} - \ln P_{\geq s})/\ln 2$ is  the local slope of a plot of $\ln P_{\geq 2s}$ vs $\ln s$, and $\red{B_3}  = B_1(2^{-\Omega}-1)/\ln 2$. Equation (\ref{localslope}) implies that if we make of plot of the local slope vs $s^{-\Omega}$ at $p_c$, linear behavior will be seen for large $s$, and the intercept \red{$s^{-\Omega}\to 0$} of the straight line will give the value of $(2-\tau)$.  Again, if we are not at $p_c$, the behavior will not be linear for large $s$.


\section{Simulation results}\label{sec:simdet}
We carried out numerical simulations using the single-cluster growth algorithm. First, a site on the lattice is chosen as the seed. Under periodic boundary conditions, any site on the lattice can be chosen as the seed. Then, an individual cluster is grown at that seeded site. To grow the clusters, we check all neighbors of a growth site for unvisited sites, which we occupy with probability $p$, or leave unoccupied with probability $1-p$, and put the newly occupied growth sites on a first-in, first-out queue. To simulate bond percolation, we simply leave the sites in the unvisited state when we do not occupy them, i.e., when rnd$<p$, where rnd is a uniformly distributed random number in $(0,1)$.  (For site percolation, sites are blocked from ever being occupied in the future, once they have been visited by the growth process.) The single-cluster growth method is similar to the Leath method \cite{Leath1976}.  A more detailed description of our algorithm is given in Ref.\ \cite{XunZiff2019}.

Some clusters will be small, while others may be very large. To keep the clusters from exceeding the system size, an upper size cutoff is set. Clusters that are smaller than the upper size cutoff can grow until they terminate in a complete cluster. For clusters larger than the upper size cutoff, their growth is halted when the size of the cluster reaches the cutoff. In fact, there are many clusters that are quite small and grow very quickly. We utilize a simple programming procedure to avoid clearing out the lattice after each cluster is formed: the lattice values are started out at $0$, and for cluster $n$, any site whose value is less than $n$ is considered unoccupied. When a site is occupied in the growth of a new cluster, it is assigned the value of the cluster number $n$. The procedure  saves a great deal of time because we can use a very large lattice, and do not have to clear out the lattice after each cluster is generated.

Another advantage of the single-cluster growth method is that it is very simple to record and analyze the results \cite{LorenzZiff1998}. We attribute clusters of different sizes to different bins. Clusters whose size (number of sites) fall in a range of $(2^n, 2^{n+1}-1)$ for $n=0,1,2,\ldots$ belong to the $n$-th bin. Clusters still growing when they reach the upper size cutoff are counted in the last bin. Then, the only thing we need to record is the number of clusters in each of the bins. Thus one does not need to study properties like the intersections of crossing probabilities for different size systems or create large output files of intermediate microcanonical results to find estimates of the threshold. The cutoff is $2^{16}$ occupied sites for all the lattices in this paper, meaning that the output files here are simply the 17 values of the bins for each value of $p$.  While the method is not as efficient as the union-find method \cite{NewmanZiff2001}, which utilizes only one set of runs to simulate all values of $p$, it has the virtue that it is simple to analyze.  If one concentrates the longest runs only to the values closest to $p_c$ (determined as one goes), the net disadvantage is not that great.

We have tested this method on the sc lattice, and find $p_c = 0.2488117(5)$, $\tau = 2.18905(5)$, and $\Omega = 0.63(3)$, consistent with previous works as mentioned above.  We do not show the details of that work here.  We have also used this method to study four different four-dimensional lattices, including one with a complex neighborhood \red{in Ref.\ }\cite{XunZiff2019}.

In this paper, simulations on the sc lattice with extended neighborhoods were carried out for system size $L\times L \times L$ with $L=512$, and with helical periodic boundary conditions. $10^9$ independent samples were produced for each lattice, representing several weeks of computer time each. Then the number of clusters greater than or equal to size $s$ was found based on the data from our simulations, and  $s^{\tau-2}P_{\geq s}$ could be easily calculated.

\begin{figure}[htbp] 
\centering
\includegraphics[width=3.8in]{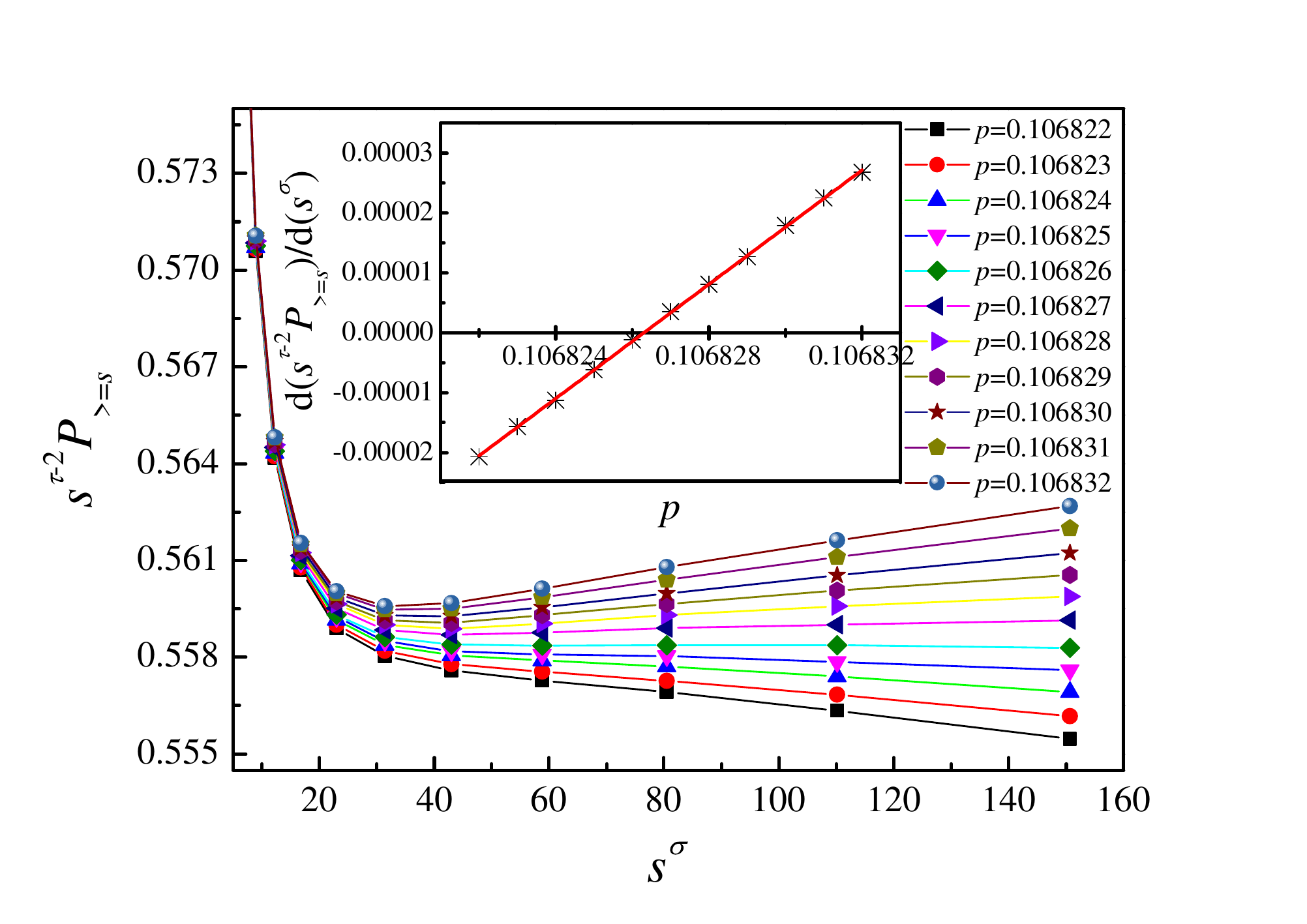} 
\caption{Plot of $s^{\tau-2}P_{\geq s}$ vs $s^{\sigma}$ with $\tau = 2.18905$ and $\sigma = 0.4522$ for the sc-NN+4NN lattice under different values of $p$. The inset indicates the slope of the linear portions of the curves shown in the main figure as a function of $p$, and the \red{threshold} value of $p_{c} = 0.1068263$ can be calculated from the $x$ intercept.}
\label{fig:sc-nn+4nn-sigma}
\end{figure}

The plot of $s^{\tau-2}P_{\geq s}$ (with $\tau = 2.18905$) vs $s^{\sigma}$ (with $\sigma = 0.4522$) for the sc-NN+4NN lattice for different values of $p$ is shown in Fig.\ \ref{fig:sc-nn+4nn-sigma}. For small clusters, there is a steep decline due to finite-size effects. For large clusters, the plot shows a linear region. The closer $p$ is to $p_{c}$, the linear portions of the curve become more nearly horizontal. Then the value of $p_{c}$ can be deduced by plotting the slope of that linear part vs $p$, since by (\ref{vssigma}),
\begin{equation}
\frac{\mathrm{d} (s^{\tau-2}P_{\geq s})}{\mathrm{d} (s^{\sigma})} \sim D_2(p-p_{c}),
\end{equation}
Finding the intercept where the derivative equals zero yields $p_c$.  This is shown in the inset of Fig.\ \ref{fig:sc-nn+4nn-sigma}. The predicted value of the percolation threshold, which is $p_{c} = 0.1068263$, corresponds to the $x$ intercept \red{in the inset plot}.

\red{If we try different values of $\tau$, we find the value of $p_c$ changes by a small amount.  If instead we use $\tau = 2.18895$, the lower end of the accepted values of $\tau$, we find $p_c = 0.1068265$, and if we take $\tau = 2.18915$, the upper end of the accepted values, we find $p_c=0.1068261$.  Thus, we can assign to $p_c = 0.1068263$ an error of just 2 in the last digit due to the uncertainty in $\tau$.}

\begin{figure}[htbp] 
\centering
\includegraphics[width=3.8in]{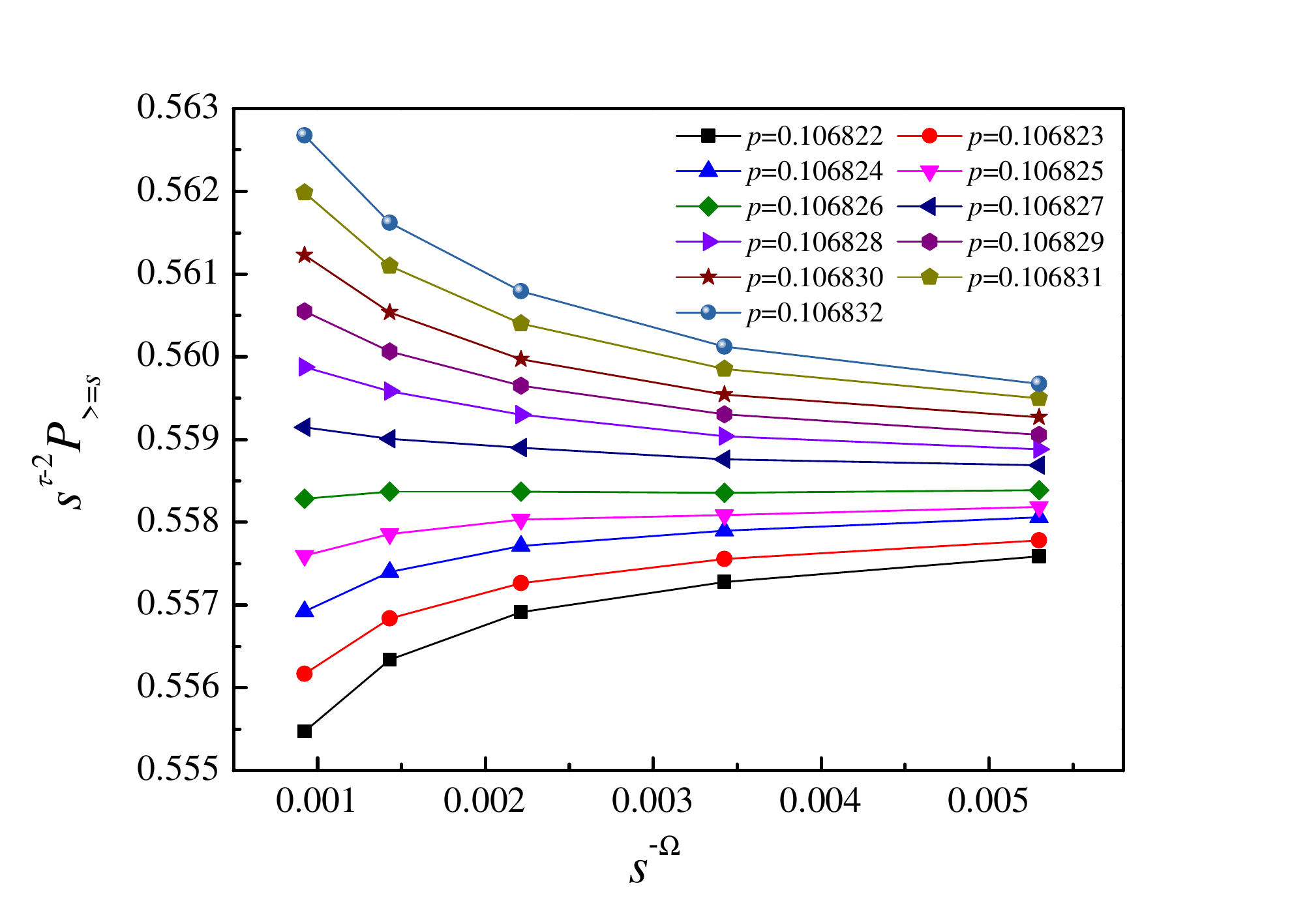} 
\caption{ Plot of $s^{\tau-2}P_{\geq s}$ vs $s^{-\Omega}$ with $\tau = 2.18905$ and $\Omega = 0.63$ for the sc-NN+4NN lattice under different values of $p$.}
\label{fig:sc-nn+4nn-omega}
\end{figure}

Fig.\ \ref{fig:sc-nn+4nn-omega} shows the plot of $s^{\tau-2}P_{\geq s}$ (with $\tau = 2.18905$) vs. $s^{-\Omega}$ (with $\Omega = 0.63$) for the sc-NN+4NN lattice under different values of $p$ for large clusters. When $p$ is very near to $p_{c}$, we can see better linear behavior, while the curves show a deviation from linearity if $p$ is away from $p_{c}$. From this plot, we can conclude that $0.106826 < p_{c} < 0.106827$, which is consistent with the value we just deduced from Fig.\ \ref{fig:sc-nn+4nn-sigma}.  

Comprehensively considering the two methods above, as well as the errors for the values of $\tau = 2.18905(15)$ and $\Omega = 0.63(4)$ (we take large error bars
for the sake of safety), we deduce the percolation threshold of the sc-NN+4NN lattice to be $p_{c} = 0.1068263(7)$, where the number in parentheses represents the estimated error in the last digit.

\begin{figure}[htbp] 
\centering
\includegraphics[width=3.8in]{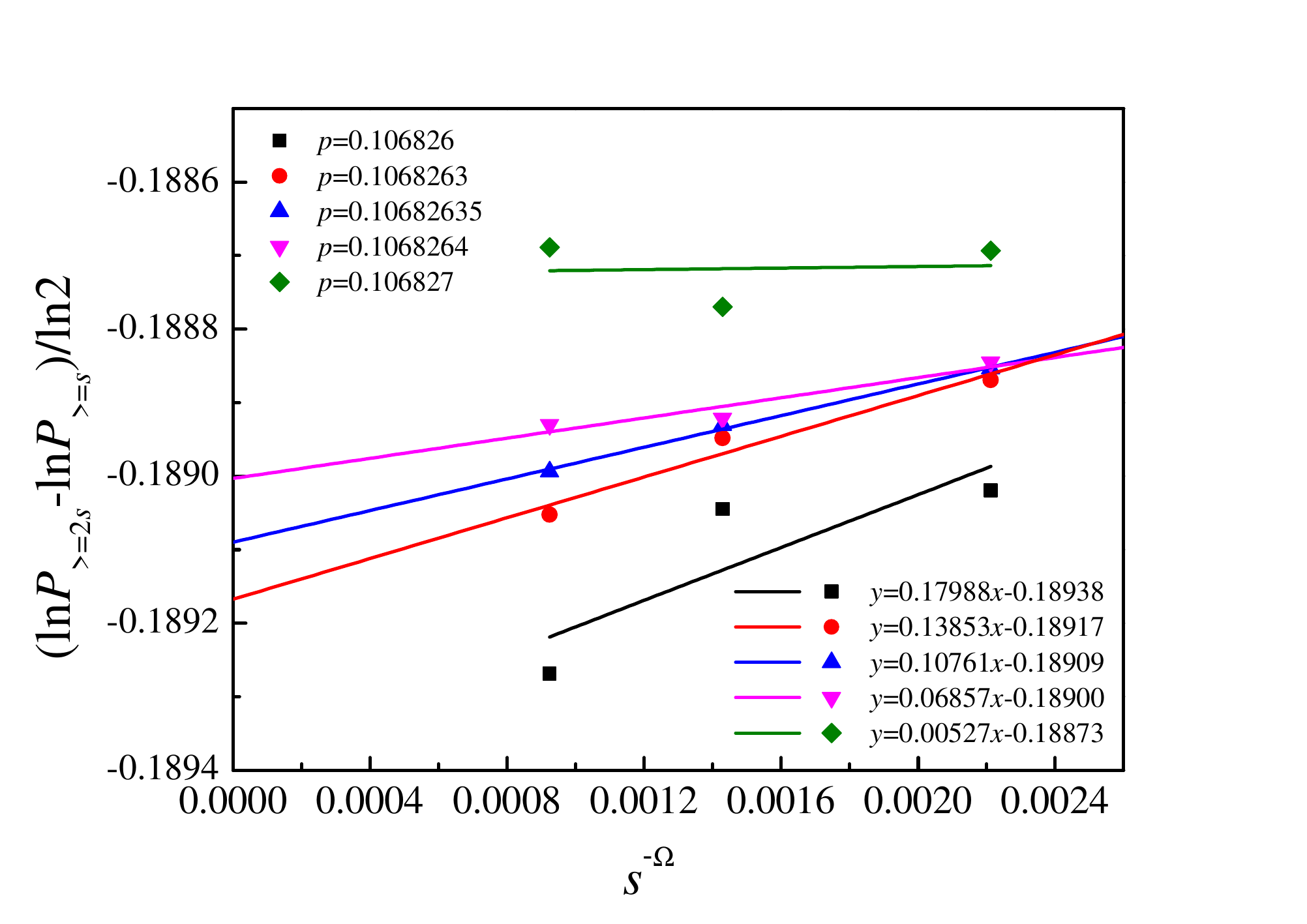} 
\caption{ Plot of local slope $(\ln P_{\geq 2s} - \ln P_{\geq s})/\ln 2$ vs $s^{-\Omega}$ with $\Omega = 0.63$ for the sc-NN+4NN lattice under values of $p = 0.106826$, $0.1068263$, $0.10682635$, $0.1068264$ and $0.106827$.}
\label{fig:sc-nn+4nn-tau}
\end{figure}

In Fig.\ \ref{fig:sc-nn+4nn-tau}, we plot the local slope (\ref{localslope}) vs $s^{-\Omega}$ with $\Omega = 0.63$ for the sc-NN+4NN lattice under the values of $p = 0.106826$, $0.1068263$, $0.10682635$, $0.1068264$, and $0.106827$. Due to the finite-size effects and the existence of longer-range bonds, we find significant higher-order corrections for smaller clusters in this lattice, and use just the last three bins for each $p$ to calculate $\tau$. We determine the value of $\tau$ falls in the interval of $(2.18900, 2.18917)$ ($\tau = 2.18900$, $2.18909$, and $2.18917$, for $p = 0.1068263$, $0.10682635$, and $0.1068264$, respectively), which is consistent with the value we use to determine $p_{c}$.

\begin{figure}[htbp] 
\centering
\includegraphics[width=3.8in]{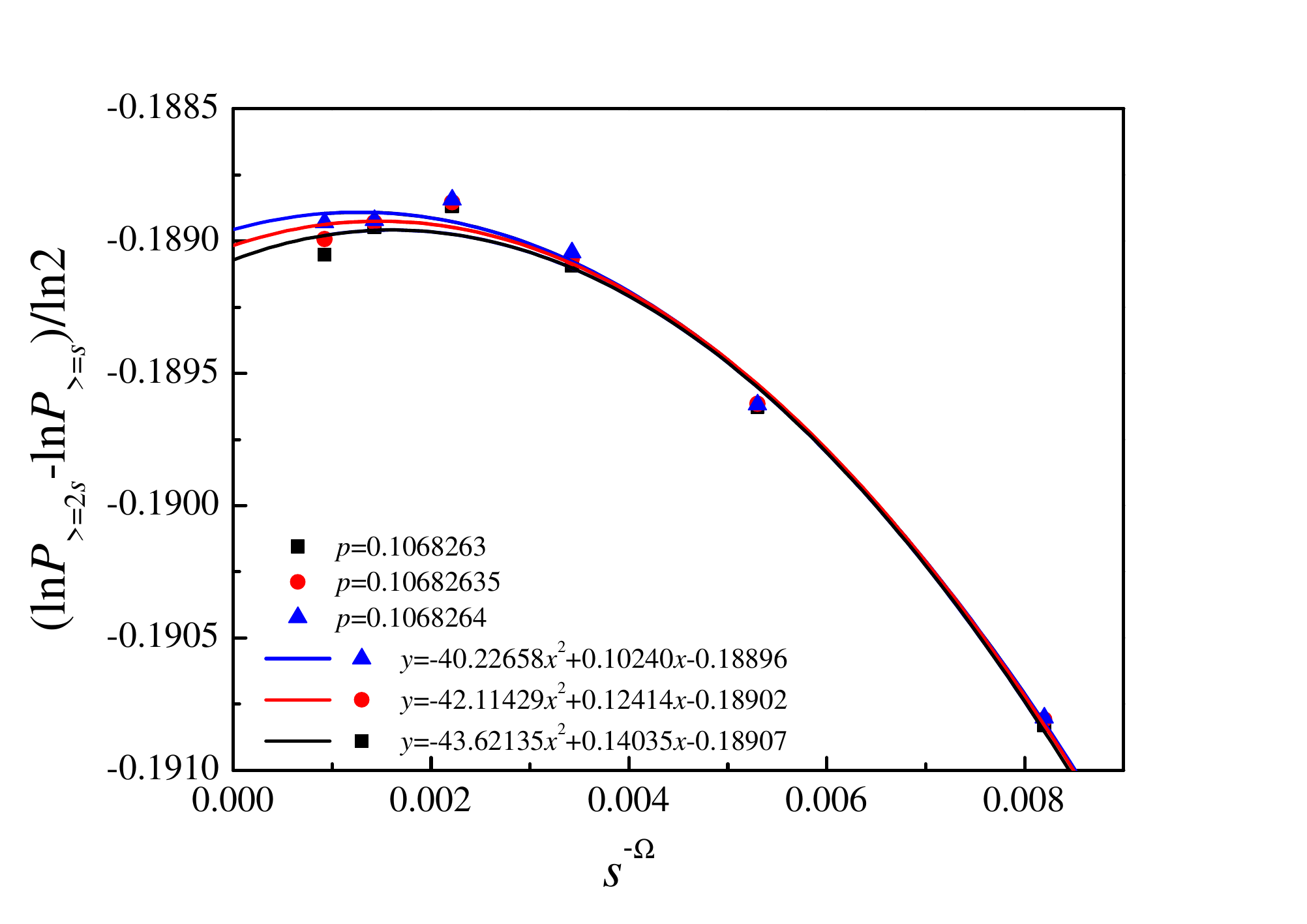} 
\caption{ Plot of local slope $(\ln P_{\ge 2s}-\ln P_{\ge s})/\ln 2$ vs $s^{-\Omega}$ for the sc-NN+4NN lattice under different values of $p$, considering second-order finite-size corrections.}
\label{fig:sc-nn+4nn-tau2}
\end{figure}

If we plotted points representing slopes from the last six bins, for example, we would have to use a quadratic to fit the data, as shown in Fig.\ \ref{fig:sc-nn+4nn-tau2}.  Here we are effectively assuming the next-order correction has exponent $2 \Omega$.  However, the fit is not that good and the intercept does not agree with the value of $\tau$ found above, so we do not consider higher-order corrections further.  In fact, we do not report any of the plots of the local  slopes (\ref{localslope}) for the other lattices.


The simulation results for the other ten lattices we considered
are shown in the Supplementary Material in Figs.\ 1-20.
and the corresponding percolation thresholds are summarized in Table\ \ref{tab:perholds}. We did not calculate the values of $\tau$ for all these lattices one by one; otherwise, the overall simulation time would at least double.  For all these plots we assumed the values $\tau = 2.18905$ that we found for the sc lattice, and $\sigma = 0.4522$.

\begin{table}[htb]
\caption{ Bond percolation thresholds determined here for the simple cubic (sc) lattice with combinations of nearest-neighbors (NN), second nearest-neighbors (2NN), third nearest-neighbors (3NN), and fourth nearest-neighbors (4NN).  \red{Also shown for reference are the site thresholds from $^a$Ref.\ \cite{Malarz2015}, $^b$Ref.\ \cite{KurzawskiMalarz2012}}}
\begin{tabular}{c|c|c|c}
\hline\hline
    lattice           & $z$ & $p_{c}$(bond)   & \red{$p_c$(site) }    \\ \hline
    sc-NN+4NN         & 12  & 0.1068263(7)  &  0.15040(12)$^a$\\
    sc-3NN+4NN        & 14  & 0.1012133(7) & 0.20490(12)$^a$  \\ 
    sc-NN+3NN         & 14  & 0.0920213(7)  & 0.1420(1)$^b$\\
    sc-NN+2NN         & 18  & 0.0752326(6)  & 0.1372(1)$^b$\\
    sc-2NN+4NN        & 18  & 0.0751589(9) & 0.15950(12)$^a$ \\
    sc-2NN+3NN        & 20  & 0.0629283(7)   & 0.1036(1)$^b$\\
    sc-NN+3NN+4NN     & 20  & 0.0624379(9) & 0.11920(12)$^a$ \\
    sc-NN+2NN+4NN     & 24  & 0.0533056(6) &  0.11440(12)$^a$\\
    sc-NN+2NN+3NN     & 26  & 0.0497080(10) &  0.0976(1)$^b$\\
    sc-2NN+3NN+4NN    & 26  & 0.0474609(9)  & 0.11330(12)$^a$\\
    sc-NN+2NN+3NN+4NN & 32  & 0.0392312(8) &  0.10000(12)$^a$\\
\hline\hline
\end{tabular}
\label{tab:perholds}
\end{table}

\section{Discussion}

In Table \ref{tab:perholds}, the lattices are arranged in the order of increasing coordination number $z$. As one would expect, the values of $p_{c}$ decrease with increasing $z$.  \red{ For reference, we have also added the site percolation thresholds found in Refs.\ \cite{Malarz2015} and \cite{KurzawskiMalarz2012}.  Note that the ordering of the site thresholds is not always the same as for the bond thresholds, and  the site thresholds are not all monotonic with $z$ as the bond thresholds are.}

\begin{figure}[htbp] 
\centering
\includegraphics[width=3.8in]{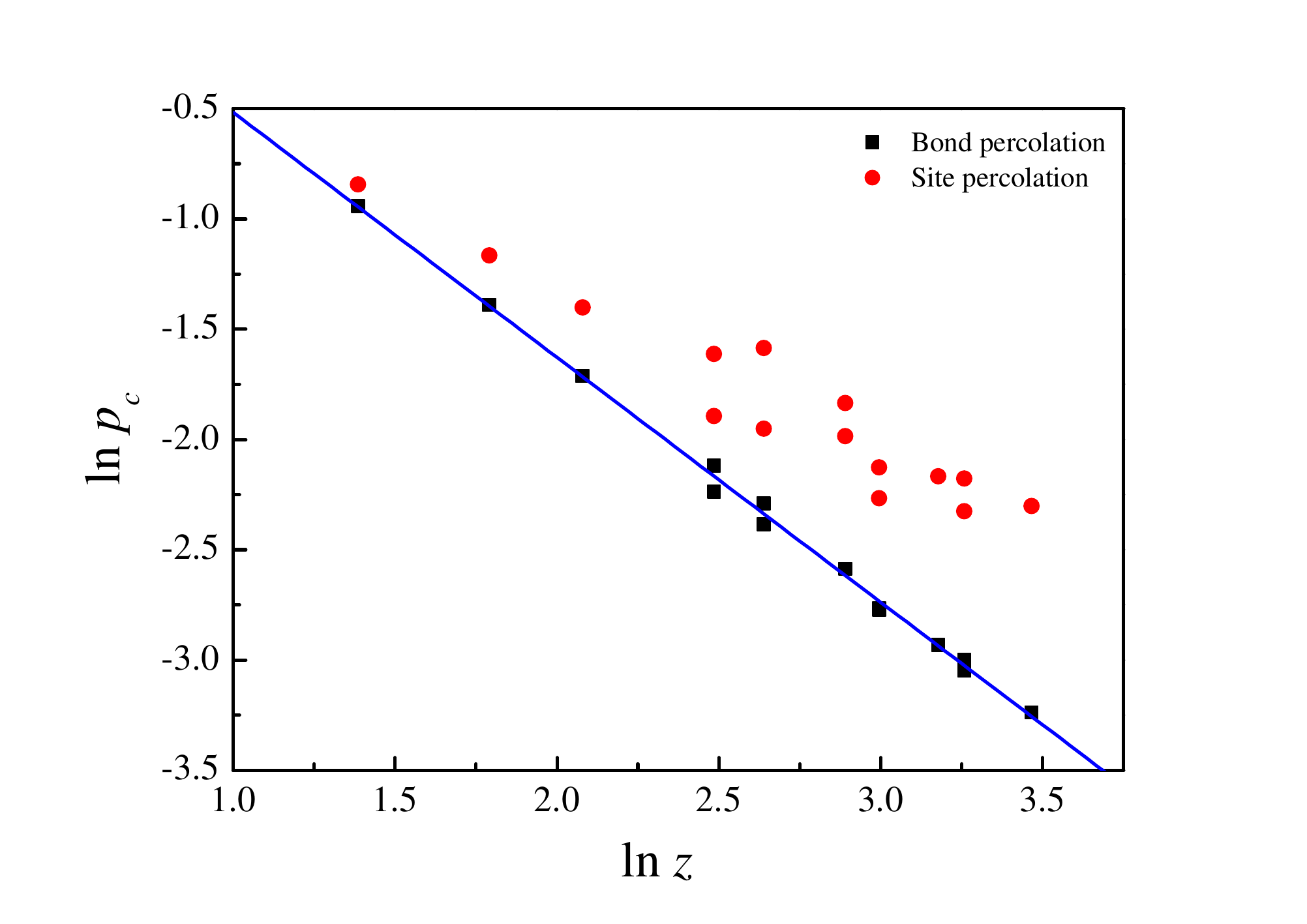}
\caption{ A log-log plot of percolation thresholds $p_{c}$ vs coordination number $z$ \red{(squares)} for the diamond lattice, the sc lattice, the bcc lattice, the fcc lattice, and the lattices simulated in this paper \red{in the order of Table \ref{tab:perholds}, left to right}. The slope gives an exponent of $a = 1.111$ in Eq.\ (\ref{eq:scaling}), and the intercept ($z=1$) of the line is at $\ln p_c = 0.594$, \red{yielding the formula
$p_c \approx 1.811 z^{-1.111}$}. Also shown on the plot are the site thresholds (provided by Refs.\ \cite{XuWangLvDeng2014,LorenzZiff1998,KurzawskiMalarz2012,Malarz2015}) for the same lattices, in which case the \red{correlation of the thresholds with $z$ is not nearly as good (circles).}}
\label{fig:ln-pc-z-gamma}
\end{figure}

In percolation research, there has been a long history of studying correlations between percolation thresholds and lattice properties \cite{ScherZallen70,vanderMarck97,Wierman02,WiermanNaor05}. For example, in Ref.\ \cite{KurzawskiMalarz2012}, Kurzawski and Malarz found that the site thresholds for several three-dimensional lattices can be fitted fairly well by a simple power-law in $z$:
\begin{equation}
p_{c}(z) \sim c z^{-a},
\label{eq:scaling}
\end{equation}
with $a =  0.790(26)$. Similar power-law relations for various systems were studied by Galam and Mauger \cite{GalamMauger96}, van der Marck \cite{vanderMarck98}, and others, often in terms of $(z-1)^{-a}$ rather than vs $z^{-a}$. For bond percolation in four dimensions, we found $a = 1.087$ in Ref.\ \cite{XunZiff2019} (where we called the exponent $a$ as $\gamma_4$). 

Here we plot the log-log relation of $p_{c}$ vs $z$ in Fig.\ \ref{fig:ln-pc-z-gamma}, along with the bond percolation thresholds of $p_{c} = 0.3895892$ \cite{XuWangLvDeng2014}, $0.2488117$, $0.1802875$ \cite{LorenzZiff1998} and $0.1201635$ \cite{LorenzZiff1998} for the diamond ($z=4$), the sc ($z=6$), the bcc ($z=8$), and the fcc ($z=12$) lattices, respectively. In Fig.\ \ref{fig:ln-pc-z-gamma}, we also make a comparison with site percolation for the same lattices, using data from various sources \cite{percthresholdwiki}. It can be seen that bond percolation follows a much better linear behavior than site percolation, where there is more scatter in the plot. As $z$ increases, the relative difference between site and bond thresholds grows, because in site percolation, a single occupied site automatically \red{has the ability to connect} to the entire neighborhood at once, while for bond percolation only two sites are connected by an added bond.  By data fitting, we deduce $a = 1.111$ for bond percolation in three dimensions, and deviations of the thresholds from the line are within about 5\% (except $\approx$ 7\% for the sc-NN+4NN lattice). 

For site percolation, one might expect $a = 1$ for compact neighborhoods and large $z$, because such neighborhoods can represent the overlap of extended objects.  For example, consider the percolation of overlapping spheres in a continuum.   Here the percolation threshold corresponds to a total volume fraction of adsorbed spheres equal to \cite{LorenzZiff00,TorquatoJiao12},
\begin{equation}
\eta_c=\frac43\pi r^3 \frac{N}{V} \approx 0.34189
\label{eq:sphere}
\end{equation}
 where $r$ is the radius of the sphere, for $N$ particles adsorbed in a system of volume $V$.  Covering the space with a fine lattice, the system corresponds to site percolation with extended neighbors up to radius $2r$ about the central point, because two spheres of radius $r$ whose centers are separated a distance $2 r$ apart will just touch.  The ratio $N/V$ corresponds to the site occupation threshold $p_c$.  The effective $z$ is equal to the number of sites in a sphere of radius $2 r$, $z = (4/3) \pi (2 r)^3$, \red{for a simple cubic lattice}.  Then from Eq.\ (\ref{eq:sphere}) it follows that $ z p_c/8 = 0.34189$ or
 \begin{equation}
 p_c = \frac{2.73512}{z}
 \label{eq:approx}
 \end{equation}  For the site thresholds available, this gives fairly accurate estimates; for example, for site percolation on the sc-NN+2NN+3NN lattice with $z=26$, this predicts $p_c = 0.1052$, compared to the measured value of 0.0976 \cite{KurzawskiMalarz2012}.  This system is actually a cube rather than a sphere, and using the cube's continuum threshold $\eta_c = 0.32476$ \cite{TorquatoJiao12,HyytiaVirtamoLassilaOtt12}, we find even a better value of $p_c = 0.09993$. (In a future study we \red{plan to} determine site percolation thresholds with complex systems having various nearest-neighbors to test Eq.\ (\ref{eq:approx}) for higher $z$.)  In any case,  this analysis implies an exponent $a$ equal to 1, for \red{site percolation} systems with compact neighborhoods.  This argument does not seem to apply directly to bond percolation, although in general bond thresholds scale with site thresholds, and it is known that bond thresholds are always lower than site thresholds for a given lattice \cite{Grimmett1999}, so it is not surprising that the bond thresholds should follow similar $1/z$ behavior.  Of course, we are not considering just compact neighbors (like NN, NN+2NN, NN+2NN+3NN, NN+2NN+3NN+4NN) in our analysis in Fig.\ \ref{fig:ln-pc-z-gamma}, but also more sparse ones, which may also affect the \red{apparent} scaling of exponent $a$. 
 
\begin{figure}[htbp] 
\centering
\includegraphics[width=4in]{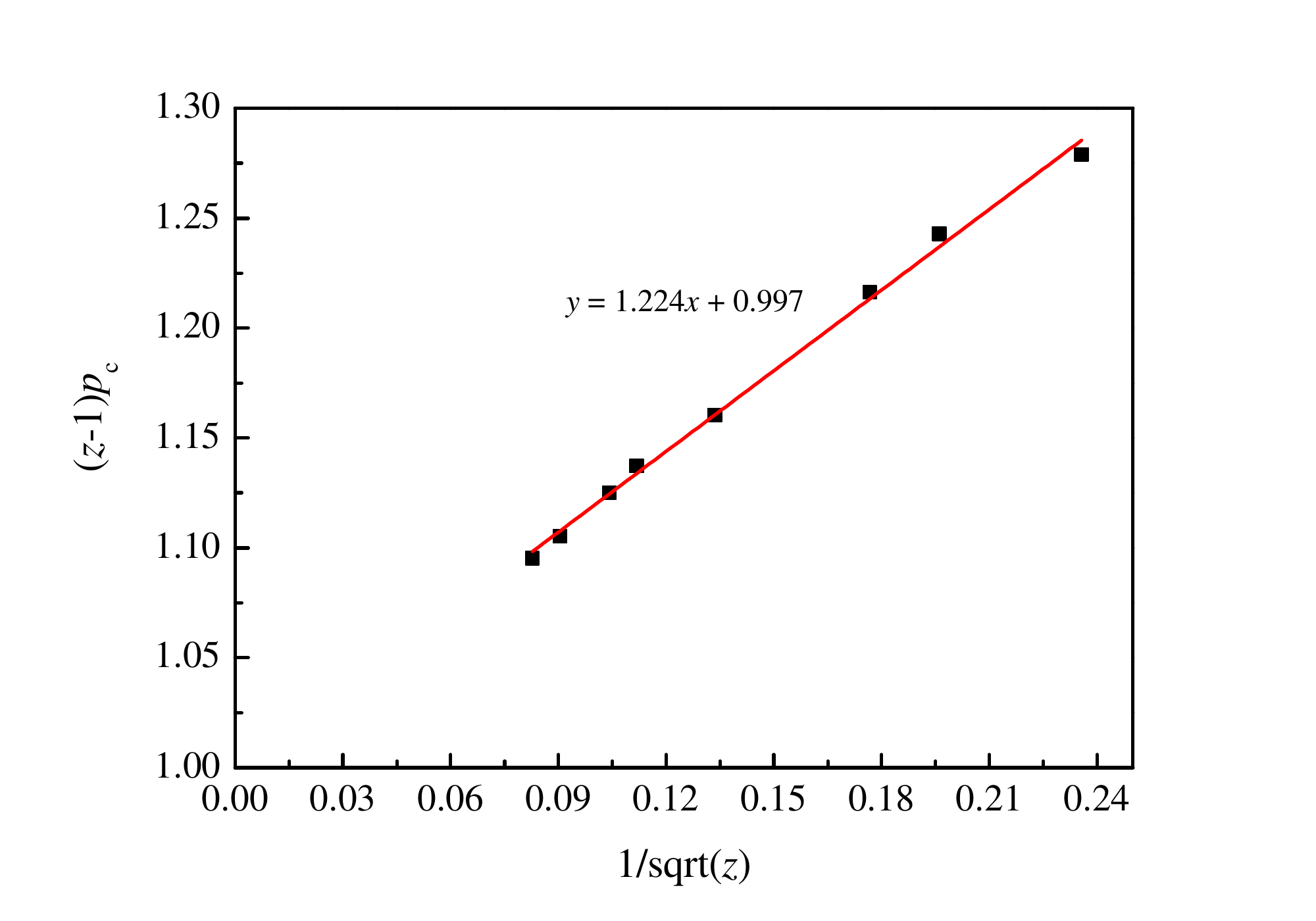}
\caption{A plot of $(z-1)p_c$ vs. $1/\sqrt{z}$ for the compact lattices sc-NN+2NN ($z=18$), +3NN ($z=26$), +4NN ($z=32$), +5NN ($z=56$), +6NN ($z=80$), +7NN ($z=92$), +8NN ($z=122$), +9NN($z=146$) (right to left), using additional threshold data for larger $z$. This plot implies the behavior shown in Eq.\ (\ref{eq:pc}).}
\label{fig:pcz1}
\end{figure}


For bond percolation, we have the bound that the threshold must be greater than that of a Bethe lattice with coordination number $z$, namely $p_c = 1/(z-1)$.  In fact, for large $z$, one would expect the Bethe result to hold asymptotically, because of the small chance that the bonds in a cluster will visit the same site.  In Fig.\ 7 we plot $(z-1)p_c$ vs $z^{-1/2}$, using additional threshold data for larger $z$, and indeed find an intercept very close to 1.  The power $-1/2$ for the correction term was found empirically.



\begin{figure}[htbp] 
\centering
\includegraphics[width=3.3in]{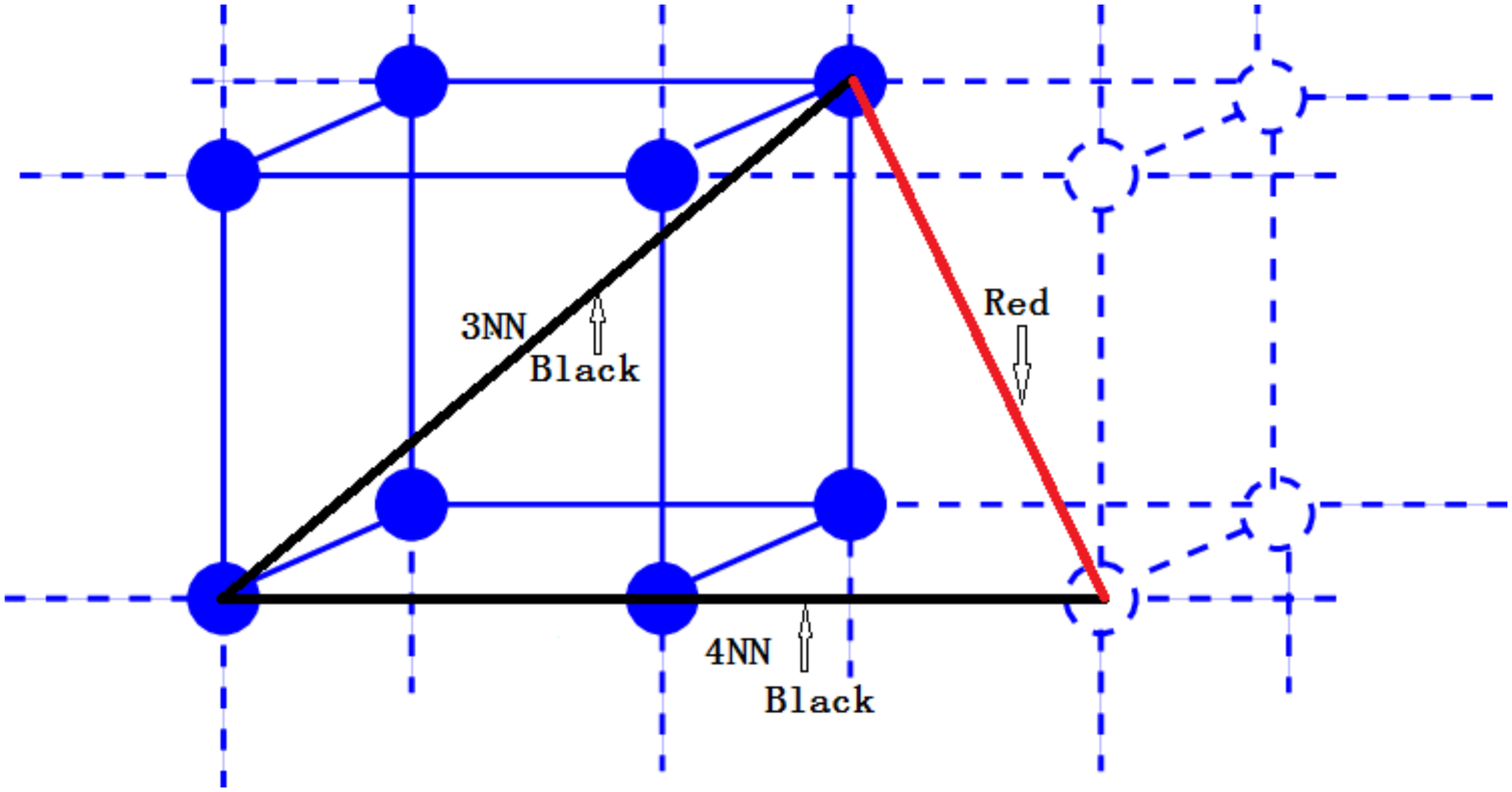}
\caption{ An example of sc-3NN+4NN cluster. Suppose the bonds in black color are occupied at the $n$-th step, then the occupation of the bond in red \red{(grey)} color will be forbidden at the ($n+1$)-th step.}
\label{fig:structure-3nn+4nn}
\end{figure}

Note that some lattices in Table\ \ref{tab:perholds} share the same $z$, but have slightly different values of $p_{c}$. For pairs of lattices with coordination number $z = 18$, $20$ and $26$, a \red{farther} distance between two neighborhood vertices seems to lead to a smaller percolation threshold. For example, for $z=18$, we have the two lattices sc-NN+2NN and sc-2NN+4NN, and the latter lattice, which has a lower percolation threshold, has 4NN vertices instead of the NN vertices of the first lattice. The exception to this trend is the sc-NN+3NN and sc-3NN+4NN lattices, both with $z = 14$, in which the latter lattice has a higher threshold.  This behavior may be due to the special cluster structure of the later lattice. An example is shown in Fig.\ \ref{fig:structure-3nn+4nn}: for the bond in red \red{(grey)} color, it is easy to form a loop, which has no contribution to percolation and, in fact, will be forbidden in our growth process \red{where we do not add bonds to previously occupied sites in the cluster.}  With the former lattice, however, loops cannot form from only three bonds, so it is easier for percolation to spread and thus the threshold is lower.  In this case the threshold is closer to the Bethe lattice prediction.

Finally, we note that for the bcc and fcc lattices with complex neighborhoods, some thresholds \red{follow} from the results of our paper here. For example, the bcc-NN+2NN lattice \red{is equivalent} to the sc-3NN+4NN lattice, and the fcc-NN+2NN  \red{lattice is equivalent to the}  sc-2NN+4NN lattices.   \red{In the same manner}, the non-complex sc-2NN lattice is equivalent to the fcc lattice, and the sc-3NN is equivalent to the bcc lattice.

\section{Conclusions}

In summary we have found precise estimates of the bond percolation threshold for eleven three-dimensional systems based upon a simple cubic lattice with multiple neighbor connections.  Similar to what we have found recently in four dimensions, the thresholds decrease monotonically with the coordination number $z$, quite accurately according to a power law of $p_{c} \sim z^{-a}$, with the exponent $a = 1.111$ here.  This compares to the value $a=1.087$ for 4d bond percolation \cite{XunZiff2019}, and the value 0.790(26) for 3d site percolation found in Ref.\ \cite{KurzawskiMalarz2012}.   However, for large $z$, the threshold must be bounded by the Bethe-lattice \red{and site percolation results}, and we find $p_c$ is given by 
\begin{equation}
p_c=\frac{1}{z-1}\left(1+1.224 z^{-1/2}\right)
\label{eq:pc}
\end{equation}
We also find that the correlation of thresholds with $z$ for bond percolation is much better than it is for site percolation.




In two, three, and higher dimensions, many percolation thresholds are still unknown, or known only to low significance, for many lattices.
Malarz and co-workers \cite{MalarzGalam05,MajewskiMalarz2007,KurzawskiMalarz2012,Malarz2015,KotwicaGronekMalarz19} have carried out several studies on lattices with various complex neighborhoods in two, three and four dimensions. Their results have all concerned site percolation, and are generally given to only three significant digits. Knowing these thresholds to higher precision, and also knowing bond thresholds, may be useful for various applications and worthy of future study.  The single-cluster algorithm is an effective way of studying these in a straightforward and efficient manner.

\section{Acknowledgments}
We are grateful to the Advanced Analysis and Computation Center of CUMT for the award of CPU hours to accomplish this work. This work is supported by the China Scholarship Council Project (Grant No. 201806425025) and the National Natural Science Foundation of China (Grant No.\ 51704293).

\bibliographystyle{unsrt}
\bibliography{bibliography.bib}

\clearpage

\appendix*

\section{Supplementary Material}

\setcounter{figure}{0}
\renewcommand{\thefigure}{A.\arabic{figure}}
Supplementary Material for ``Bond percolation on simple cubic lattices with extended neighborhoods" [Zhipeng Xun and R.\ M.\ Ziff, {\it Phys.\ Rev.\ E} 102:012102  (2020).]

\bigskip

\begin{figure}[htbp] 
\centering
\includegraphics[width=3.8in]{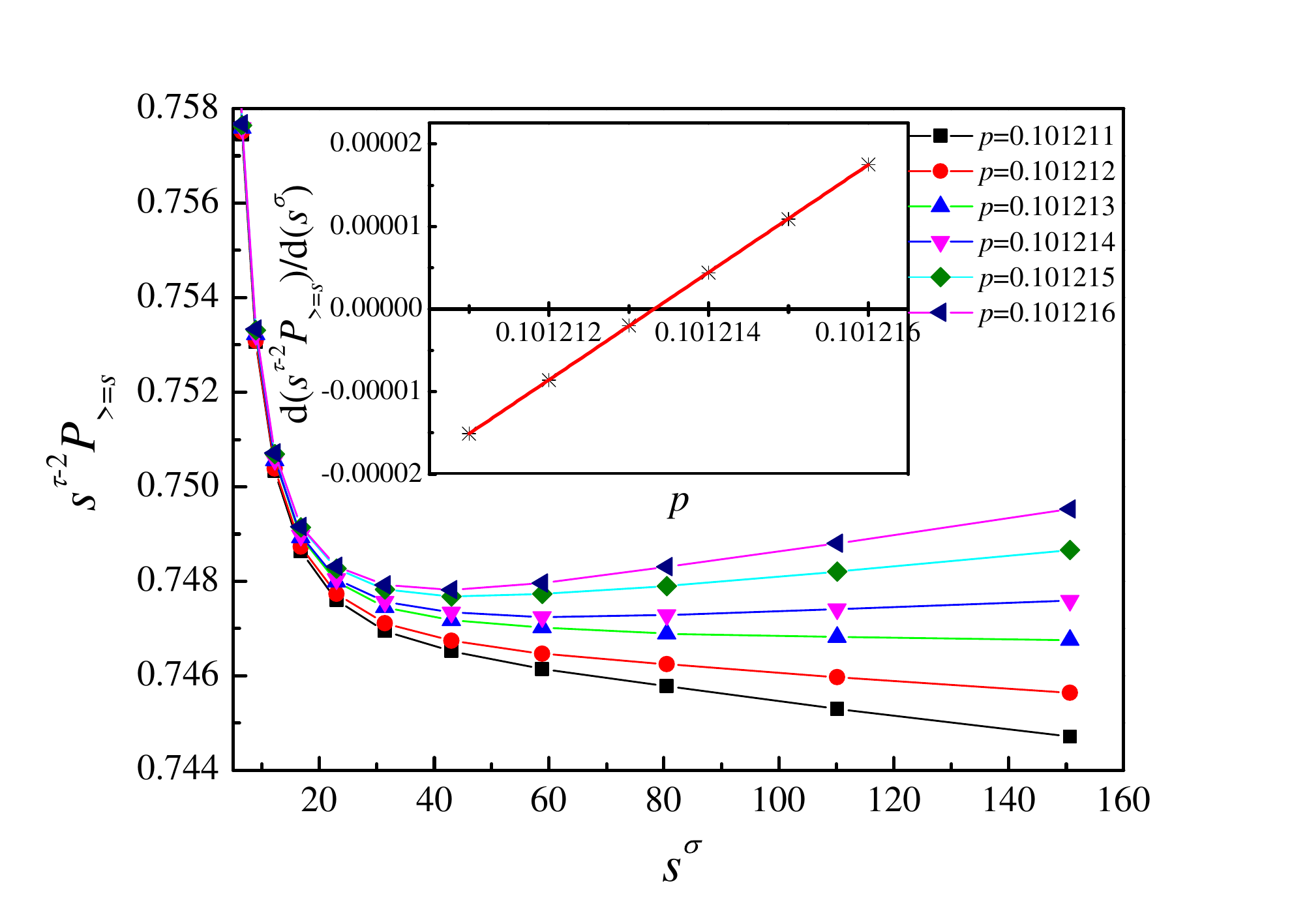} 
\caption{ Plot of $s^{\tau-2}P_{\geq s}$ vs $s^{\sigma}$ with $\tau = 2.18905$ and $\sigma = 0.4522$ for the sc-3NN+4NN lattice under different values of $p$. The inset indicates the slope of the linear portions of the curves shown in the main figure as a function of $p$, and the center value of $p_{c} = 0.1012133$ can be calculated from the $x$ intercept.}
\label{fig:sc-3nn+4nn-sigma}
\end{figure}

\begin{figure}[htbp] 
\centering
\includegraphics[width=3.8in]{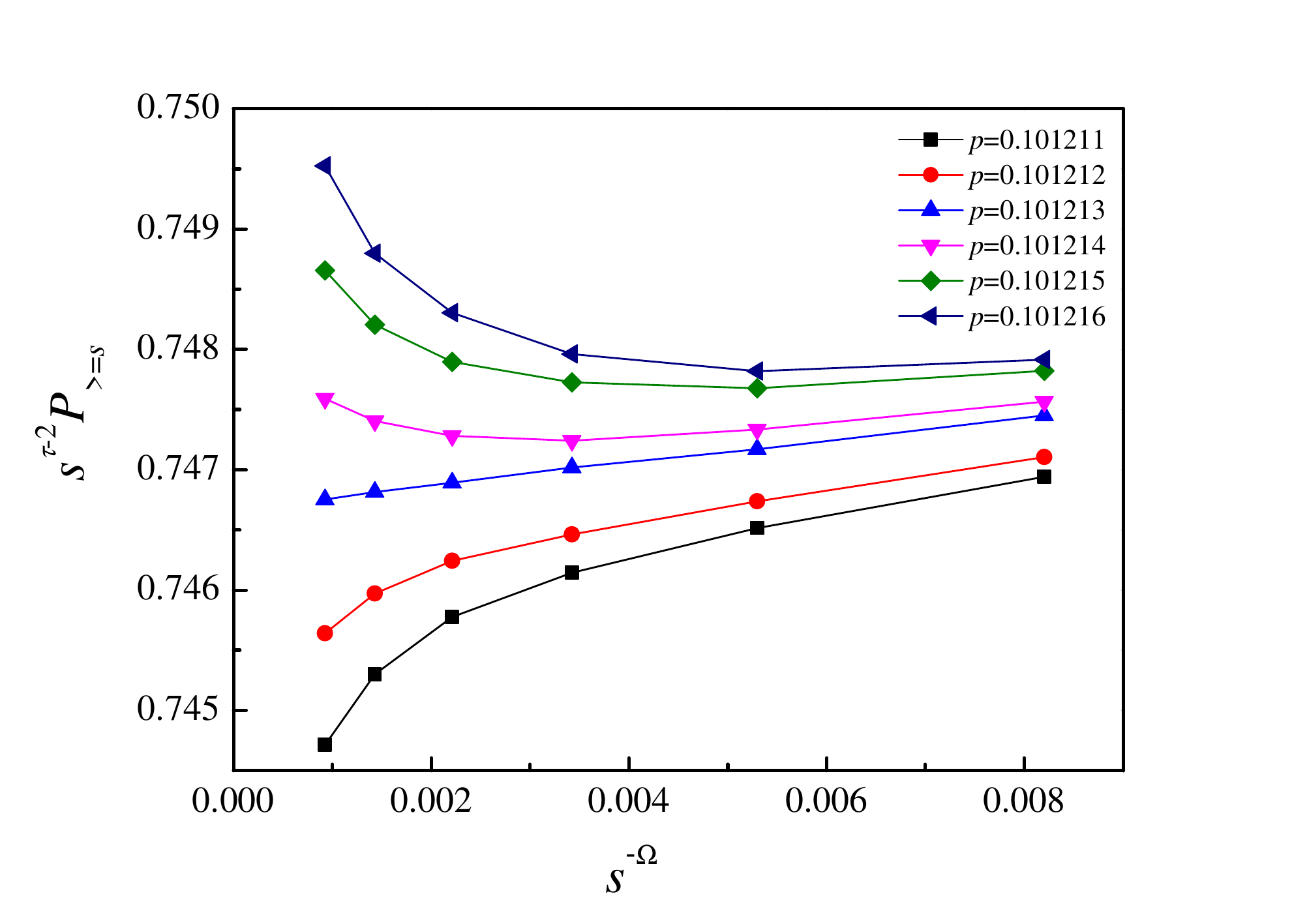} 
\caption{ Plot of $s^{\tau-2}P_{\geq s}$ vs $s^{-\Omega}$ with $\tau = 2.18905$ and $\Omega = 0.63$ for the sc-3NN+4NN lattice under different values of $p$.}
\label{fig:sc-3nn+4nn-omega}
\end{figure}

\begin{figure}[htbp] 
\centering
\includegraphics[width=3.8in]{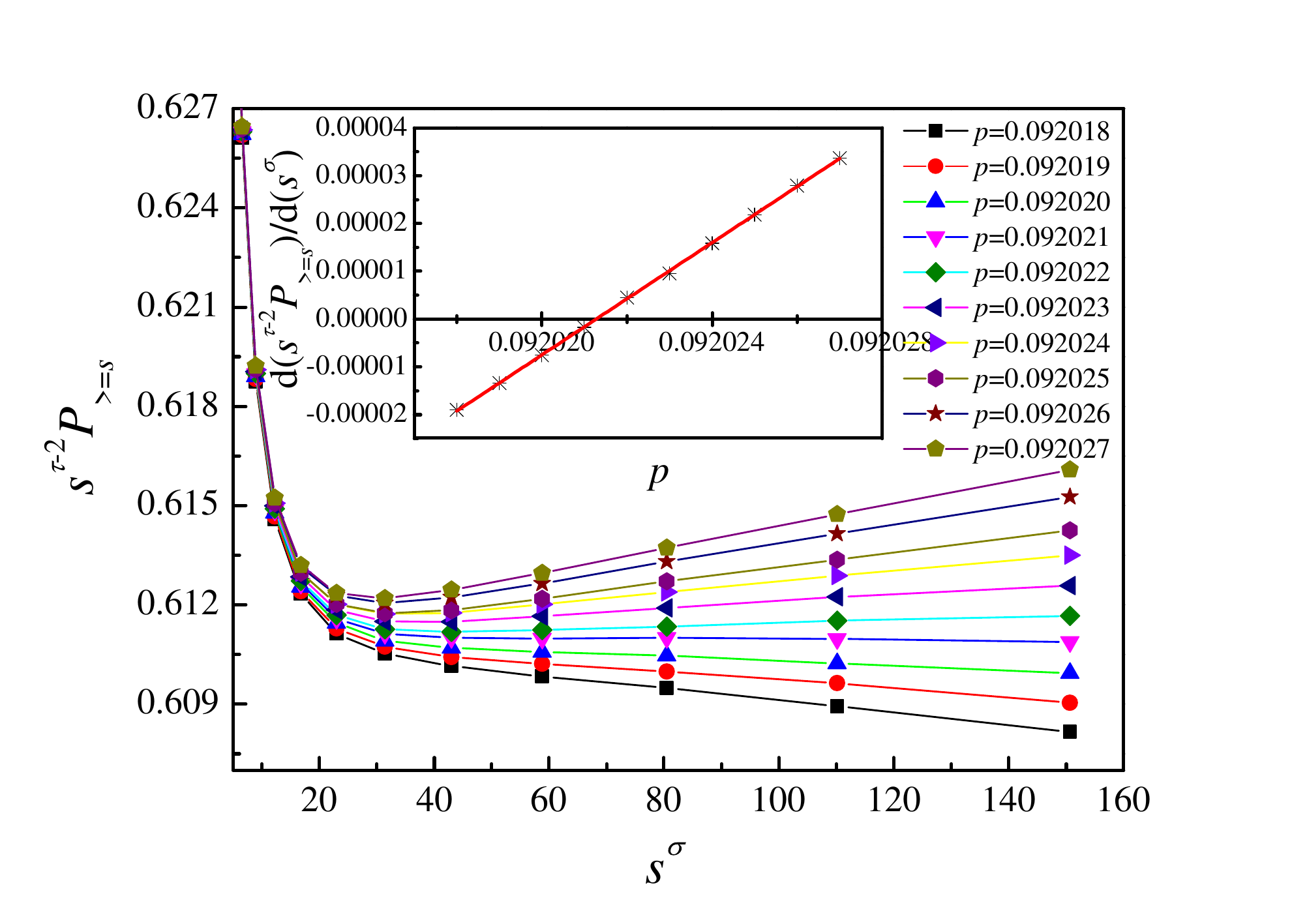} 
\caption{ Plot of $s^{\tau-2}P_{\geq s}$ vs $s^{\sigma}$ with $\tau = 2.18905$ and $\sigma = 0.4522$ for the sc-NN+3NN lattice under different values of $p$. The inset indicates the slope of the linear portions of the curves shown in the main figure as a function of $p$, and the center value of $p_{c} = 0.0920213$ can be calculated from the $x$ intercept.}
\label{fig:sc-nn+3nn-sigma}
\end{figure}

\begin{figure}[htbp] 
\centering
\includegraphics[width=3.8in]{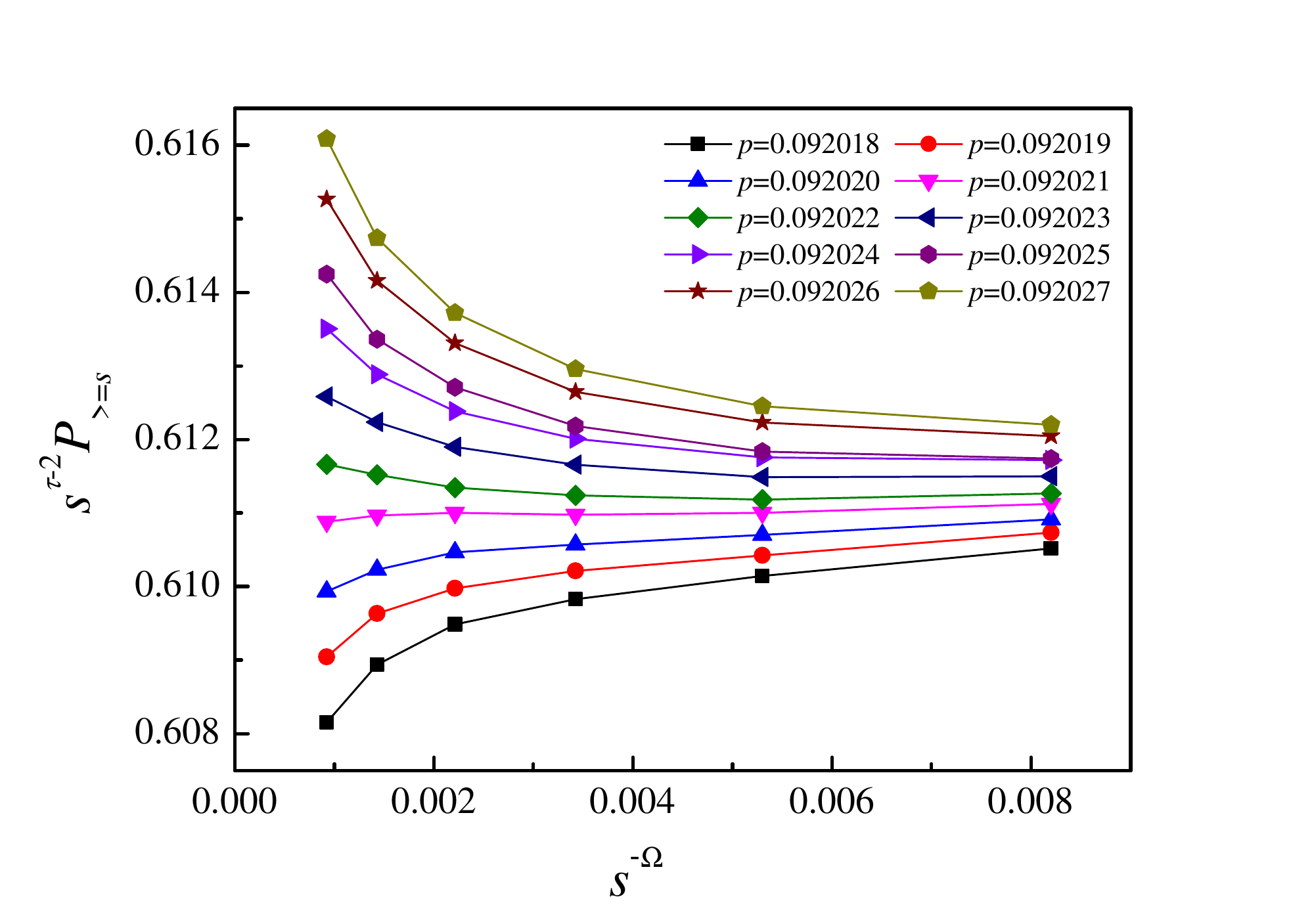} 
\caption{ Plot of $s^{\tau-2}P_{\geq s}$ vs $s^{-\Omega}$ with $\tau = 2.18905$ and $\Omega = 0.63$ for the sc-NN+3NN lattice under different values of $p$.}
\label{fig:sc-nn+3nn-omega}
\end{figure}

\begin{figure}[htbp] 
\centering
\includegraphics[width=3.8in]{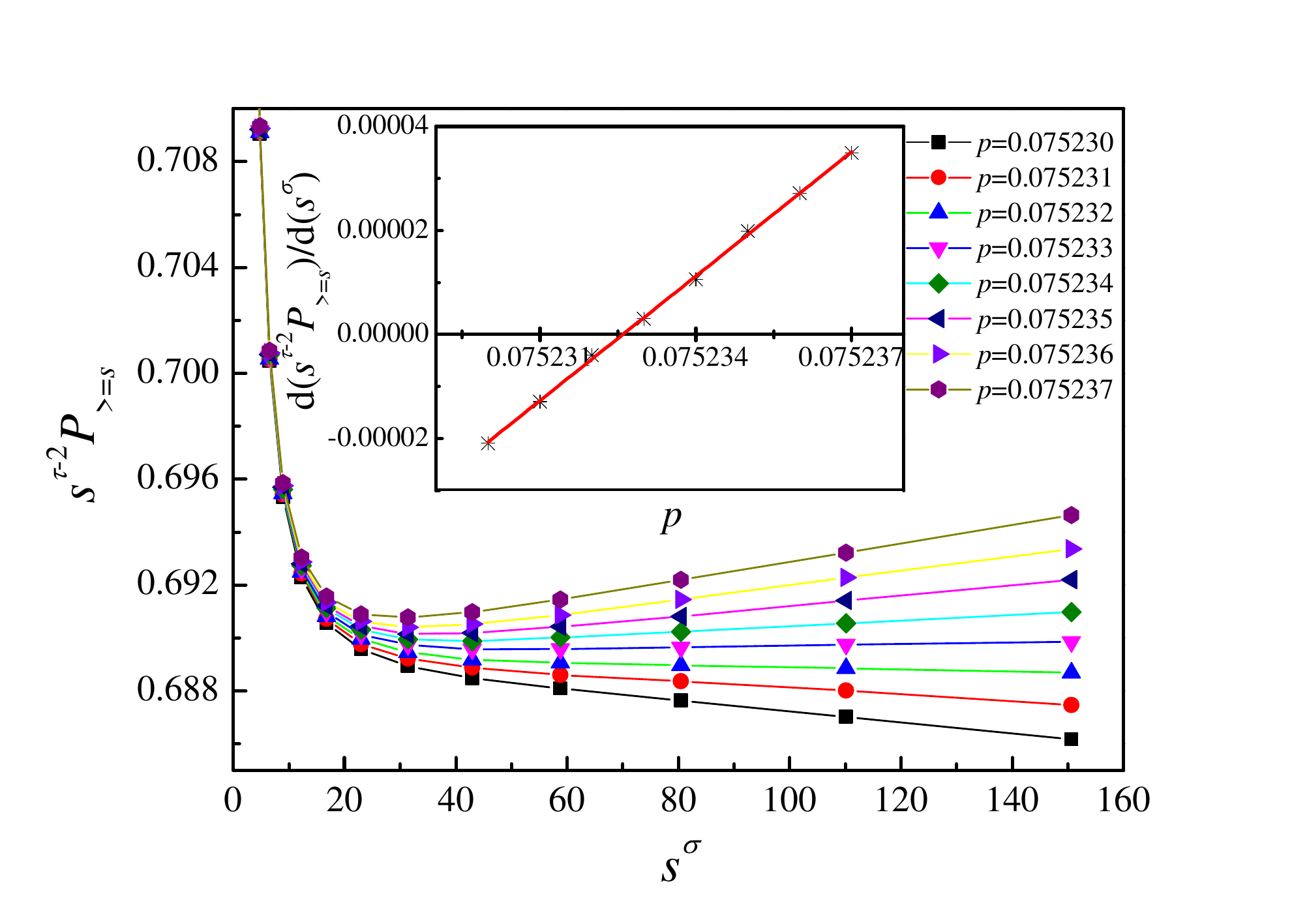} 
\caption{ Plot of $s^{\tau-2}P_{\geq s}$ vs $s^{\sigma}$ with $\tau = 2.18905$ and $\sigma = 0.4522$ for the sc-NN+2NN lattice under different values of $p$. The inset indicates the slope of the linear portions of the curves shown in the main figure as a function of $p$, and the center value of $p_{c} = 0.0752326$ can be calculated from the $x$ intercept.}
\label{fig:sc-nn+2nn-sigma}
\end{figure}

\begin{figure}[htbp] 
\centering
\includegraphics[width=3.8in]{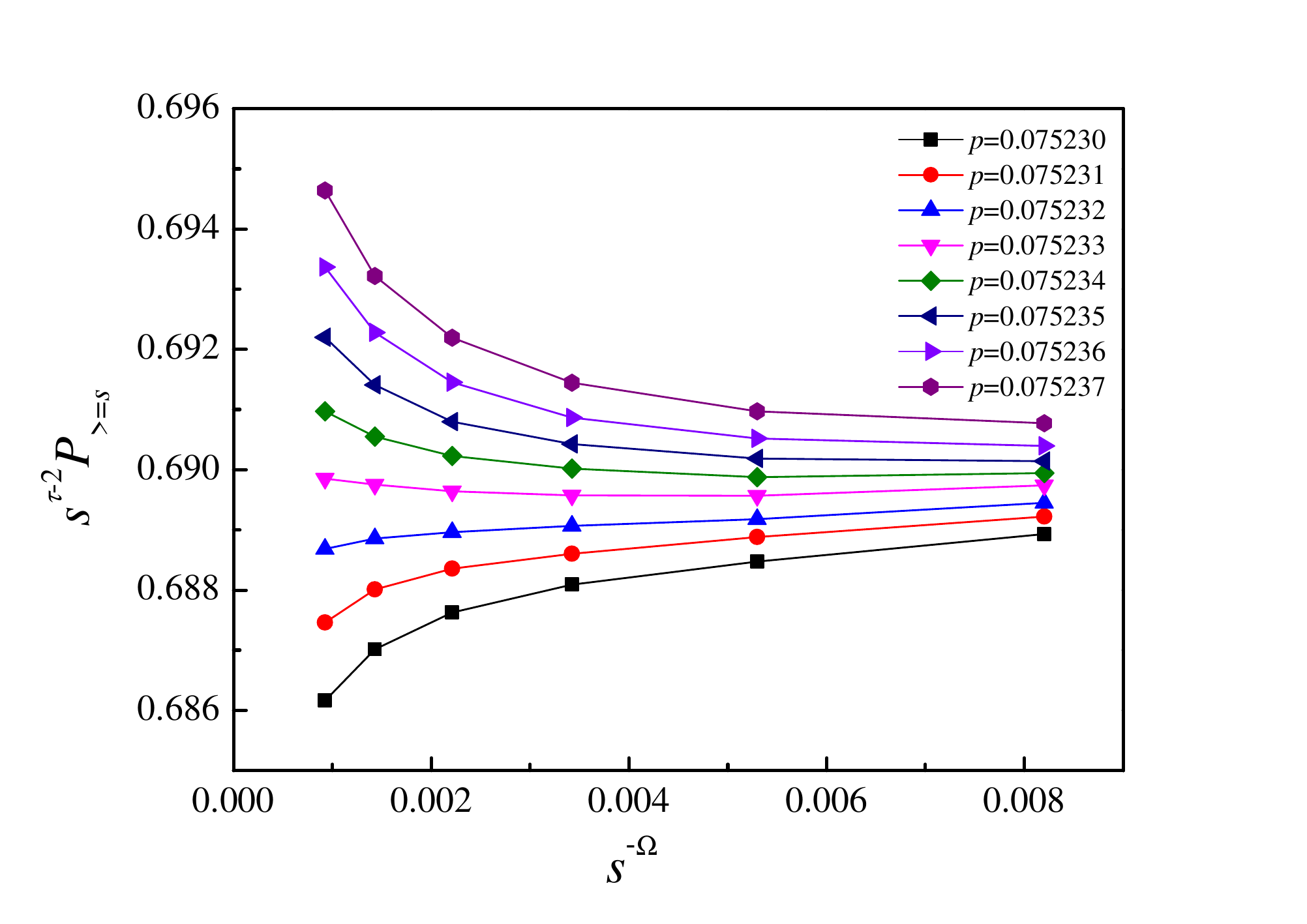} 
\caption{ Plot of $s^{\tau-2}P_{\geq s}$ vs $s^{-\Omega}$ with $\tau = 2.18905$ and $\Omega = 0.63$ for the sc-NN+2NN lattice under different values of $p$.}
\label{fig:sc-nn+2nn-omega}
\end{figure}

\begin{figure}[htbp] 
\centering
\includegraphics[width=3.8in]{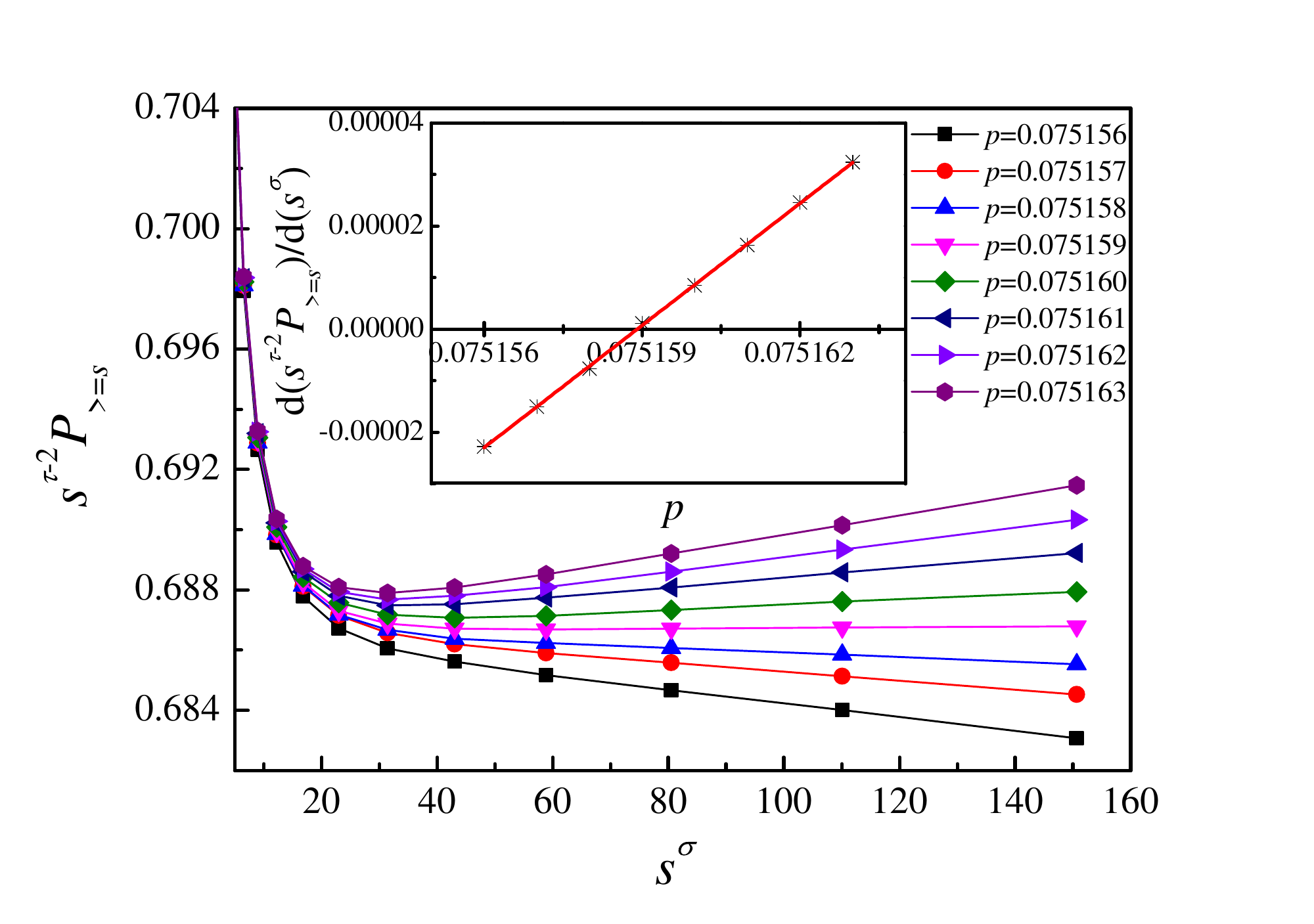} 
\caption{ Plot of $s^{\tau-2}P_{\geq s}$ vs $s^{\sigma}$ with $\tau = 2.18905$ and $\sigma = 0.4522$ for the sc-2NN+4NN lattice under different values of $p$. The inset indicates the slope of the linear portions of the curves shown in the main figure as a function of $p$, and the center value of $p_{c} = 0.0751589$ can be calculated from the $x$ intercept.}
\label{fig:sc-2nn+4nn-sigma}
\end{figure}

\begin{figure}[htbp] 
\centering
\includegraphics[width=3.8in]{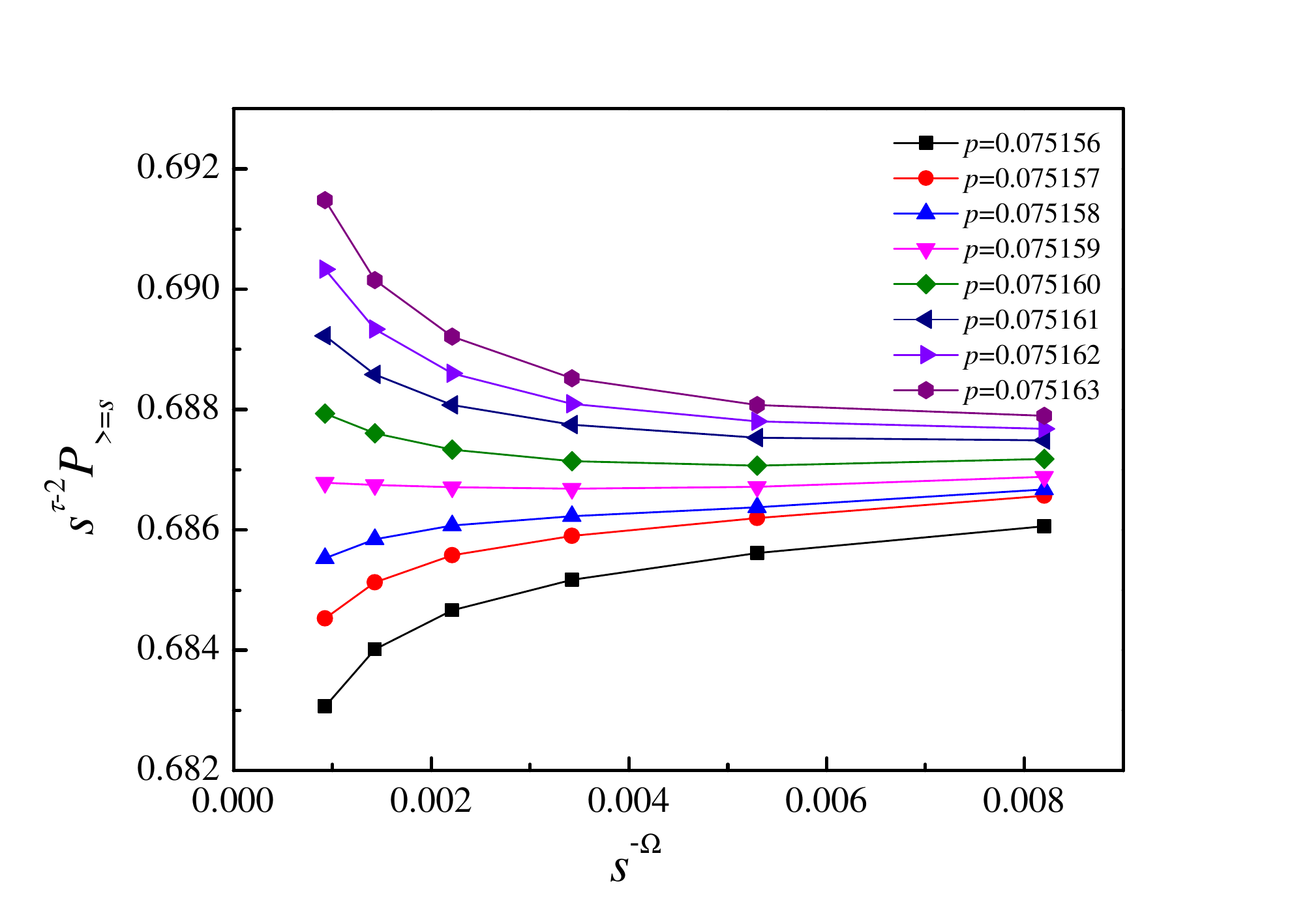} 
\caption{ Plot of $s^{\tau-2}P_{\geq s}$ vs $s^{-\Omega}$ with $\tau = 2.18905$ and $\Omega = 0.63$ for the sc-2NN+4NN lattice under different values of $p$.}
\label{fig:sc-2nn+4nn-omega}
\end{figure}

\begin{figure}[htbp] 
\centering
\includegraphics[width=3.8in]{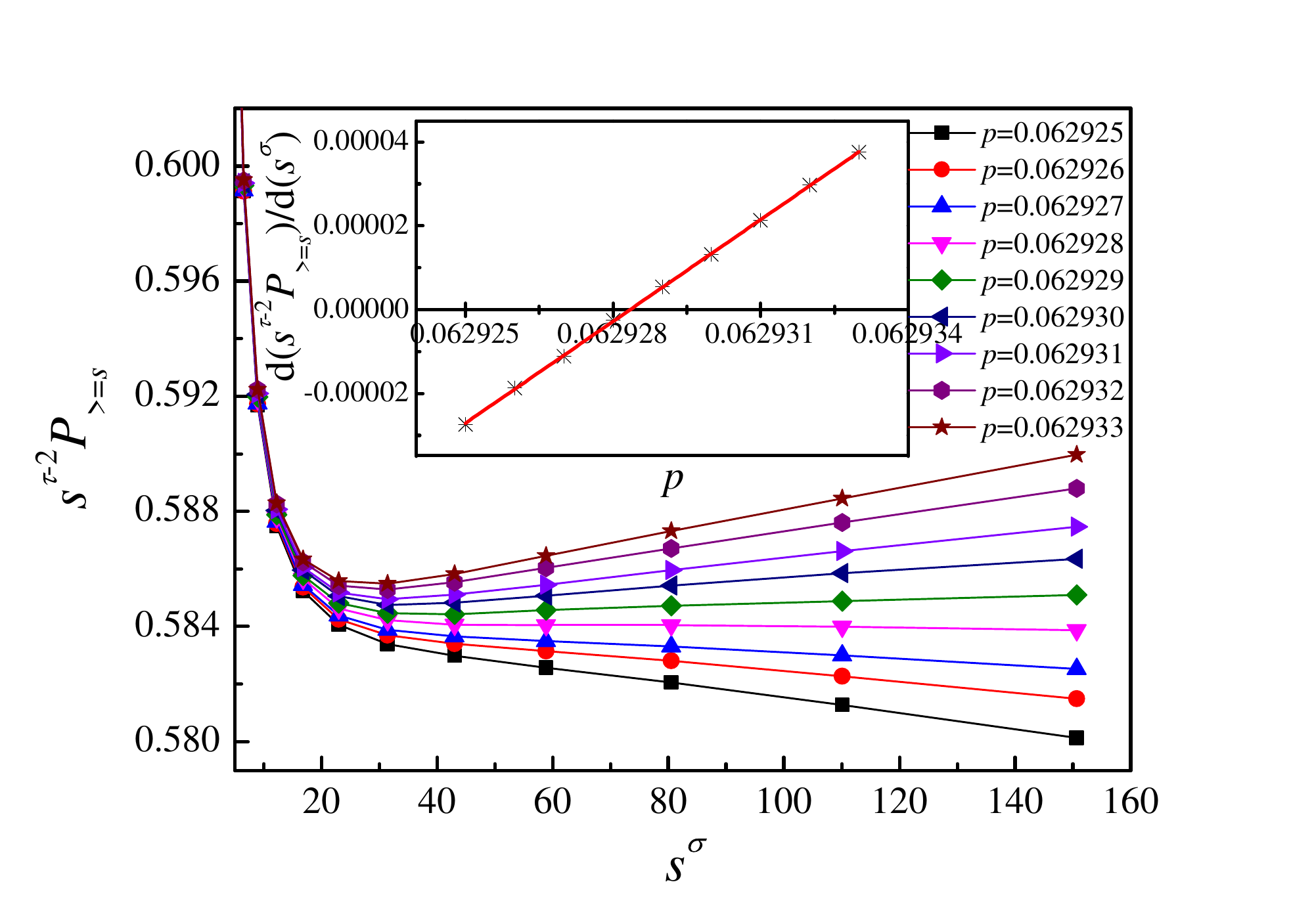} 
\caption{ Plot of $s^{\tau-2}P_{\geq s}$ vs $s^{\sigma}$ with $\tau = 2.18905$ and $\sigma = 0.4522$ for the sc-2NN+3NN lattice under different values of $p$. The inset indicates the slope of the linear portions of the curves shown in the main figure as a function of $p$, and the center value of $p_{c} = 0.0629283$ can be calculated from the $x$ intercept.}
\label{fig:sc-2nn+3nn-sigma}
\end{figure}

\begin{figure}[htbp] 
\centering
\includegraphics[width=3.8in]{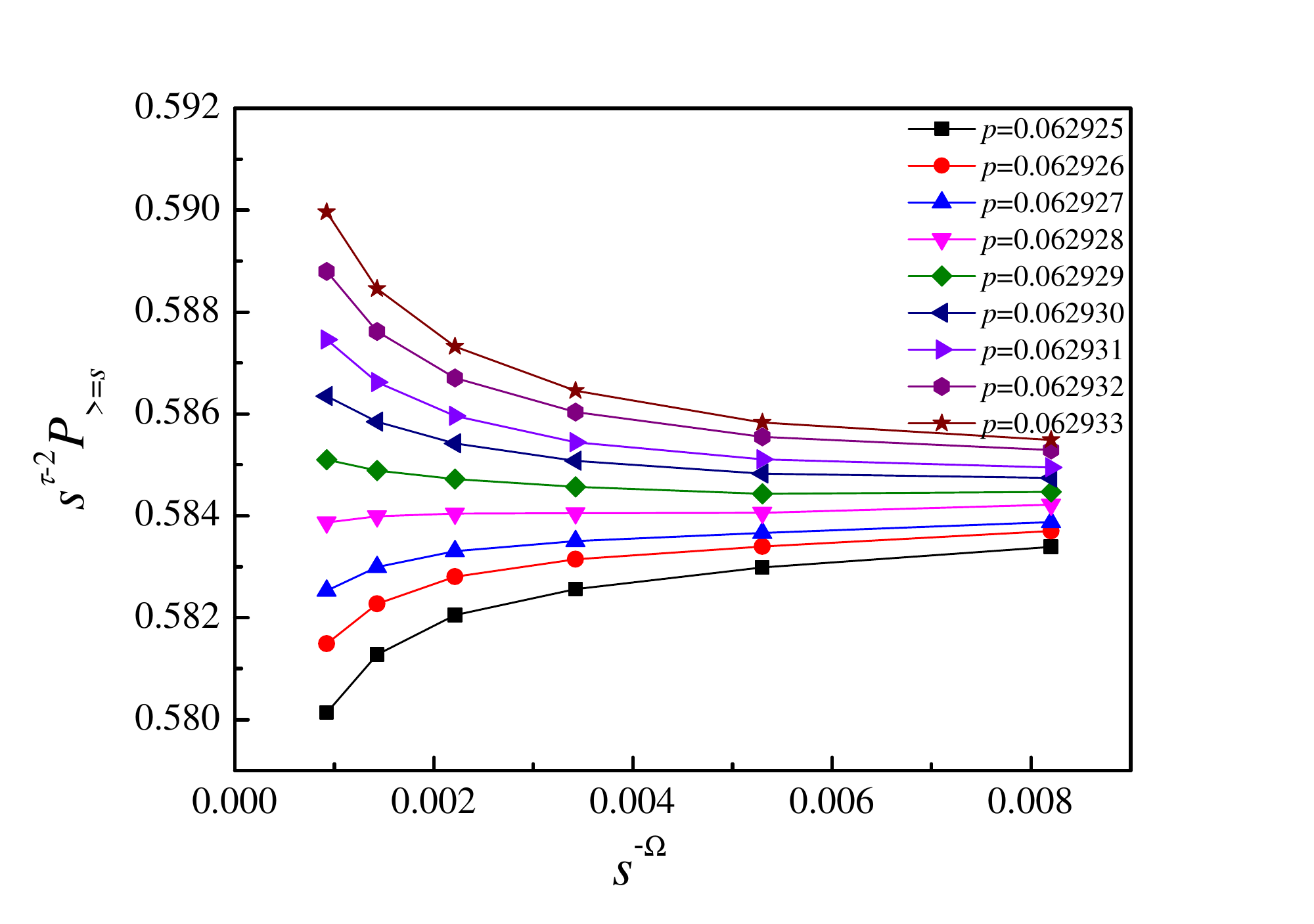} 
\caption{ Plot of $s^{\tau-2}P_{\geq s}$ vs $s^{-\Omega}$ with $\tau = 2.18905$ and $\Omega = 0.63$ for the sc-2NN+3NN lattice under different values of $p$.}
\label{fig:sc-2nn+3nn-omega}
\end{figure}

\begin{figure}[htbp] 
\centering
\includegraphics[width=3.8in]{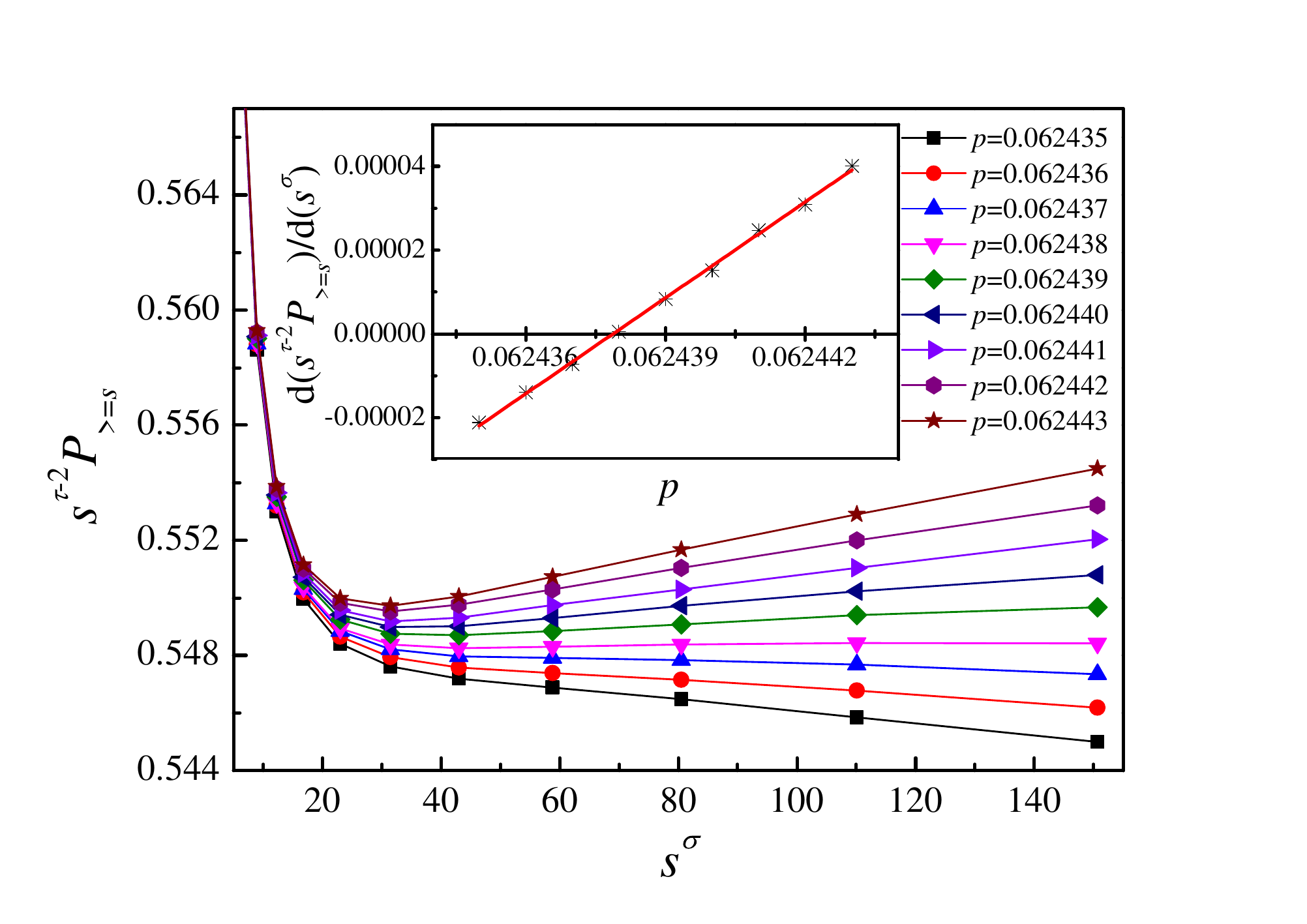} 
\caption{ Plot of $s^{\tau-2}P_{\geq s}$ vs $s^{\sigma}$ with $\tau = 2.18905$ and $\sigma = 0.4522$ for the sc-NN+3NN+4NN lattice under different values of $p$. The inset indicates the slope of the linear portions of the curves shown in the main figure as a function of $p$, and the center value of $p_{c} = 0.0624379$ can be calculated from the $x$ intercept.}
\label{fig:sc-nn+3nn+4nn-sigma}
\end{figure}

\begin{figure}[htbp] 
\centering
\includegraphics[width=3.8in]{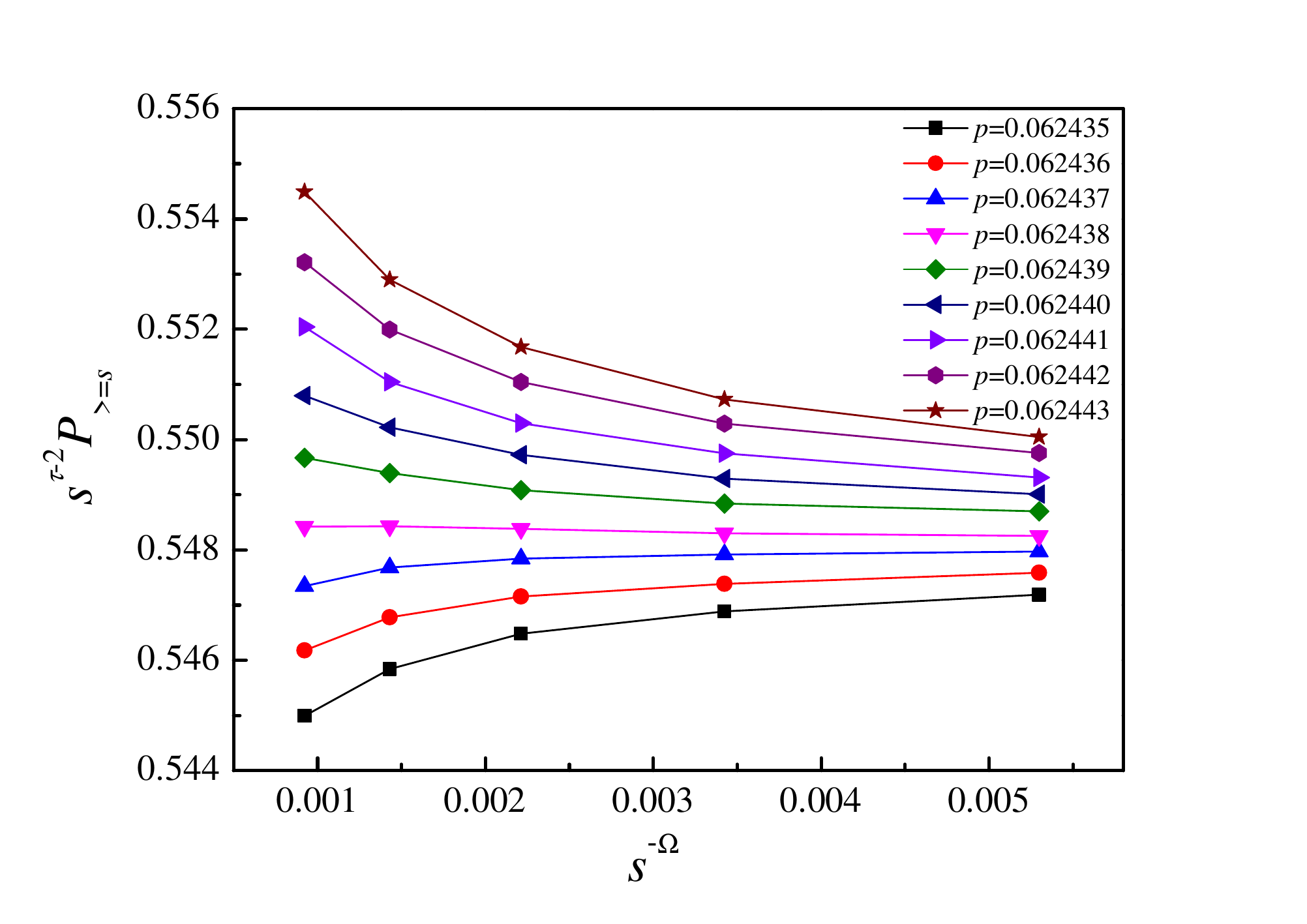} 
\caption{ Plot of $s^{\tau-2}P_{\geq s}$ vs $s^{-\Omega}$ with $\tau = 2.18905$ and $\Omega = 0.63$ for the sc-NN+3NN+4NN lattice under different values of $p$.}
\label{fig:sc-nn+3nn+4nn-omega}
\end{figure}

\begin{figure}[htbp] 
\centering
\includegraphics[width=3.8in]{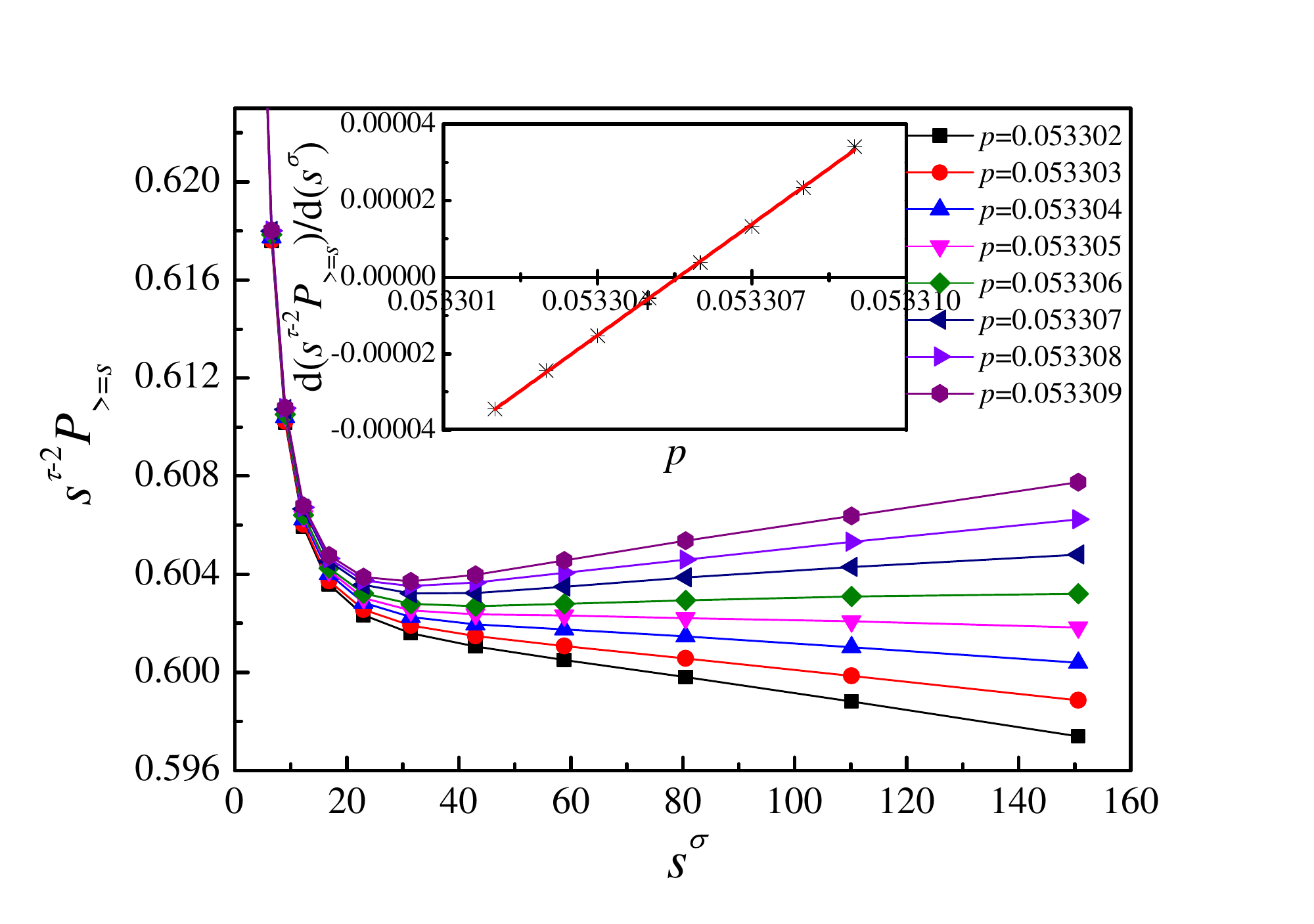} 
\caption{ Plot of $s^{\tau-2}P_{\geq s}$ vs $s^{\sigma}$ with $\tau = 2.18905$ and $\sigma = 0.4522$ for the sc-NN+2NN+4NN lattice under different values of $p$. The inset indicates the slope of the linear portions of the curves shown in the main figure as a function of $p$, and the center value of $p_{c} = 0.0533056$ can be calculated from the $x$ intercept.}
\label{fig:sc-nn+2nn+4nn-sigma}
\end{figure}

\begin{figure}[htbp] 
\centering
\includegraphics[width=3.8in]{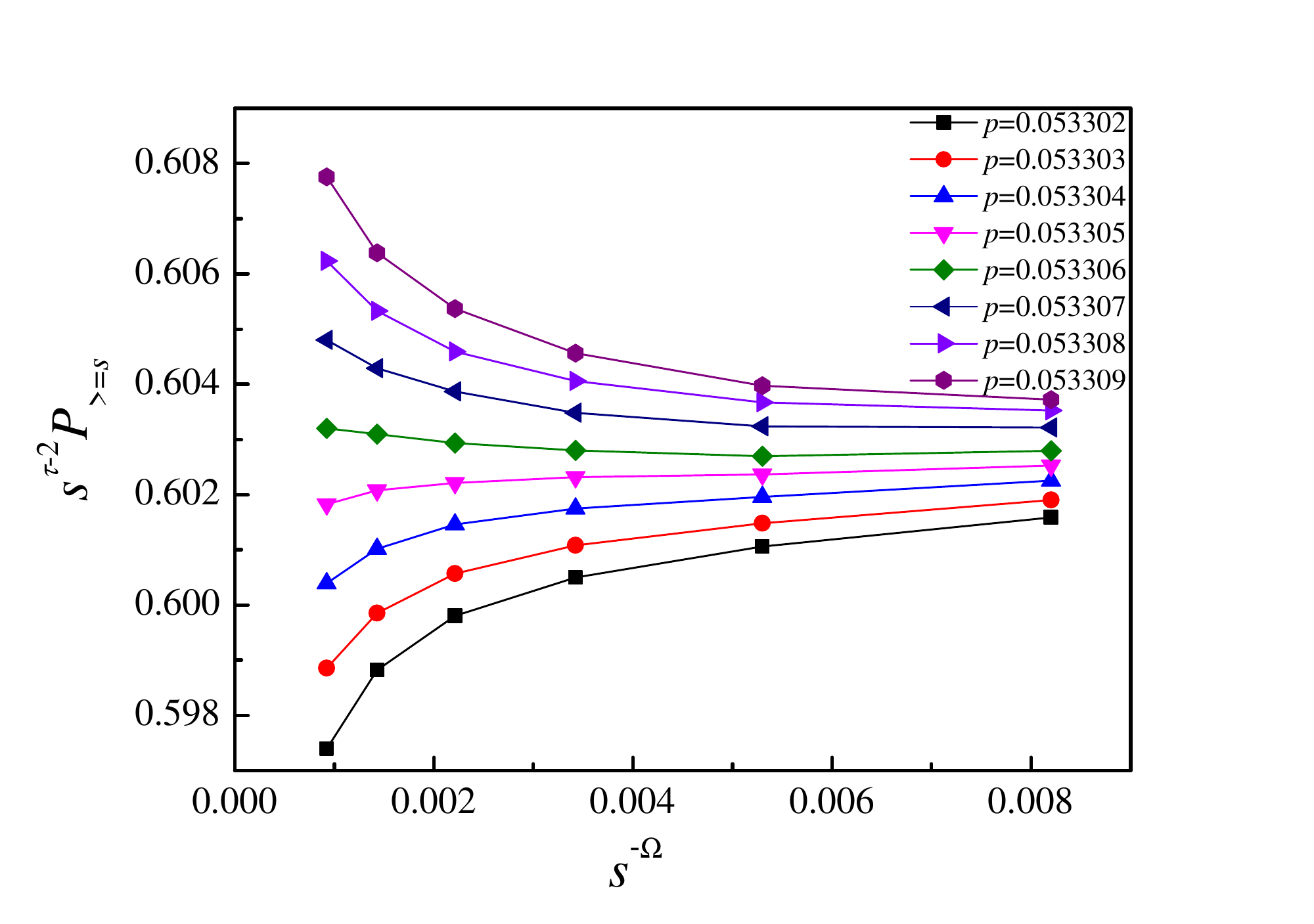} 
\caption{ Plot of $s^{\tau-2}P_{\geq s}$ vs $s^{-\Omega}$ with $\tau = 2.18905$ and $\Omega = 0.63$ for the sc-NN+2NN+4NN lattice under different values of $p$.}
\label{fig:sc-nn+2nn+4nn-omega}
\end{figure}

\begin{figure}[htbp] 
\centering
\includegraphics[width=3.8in]{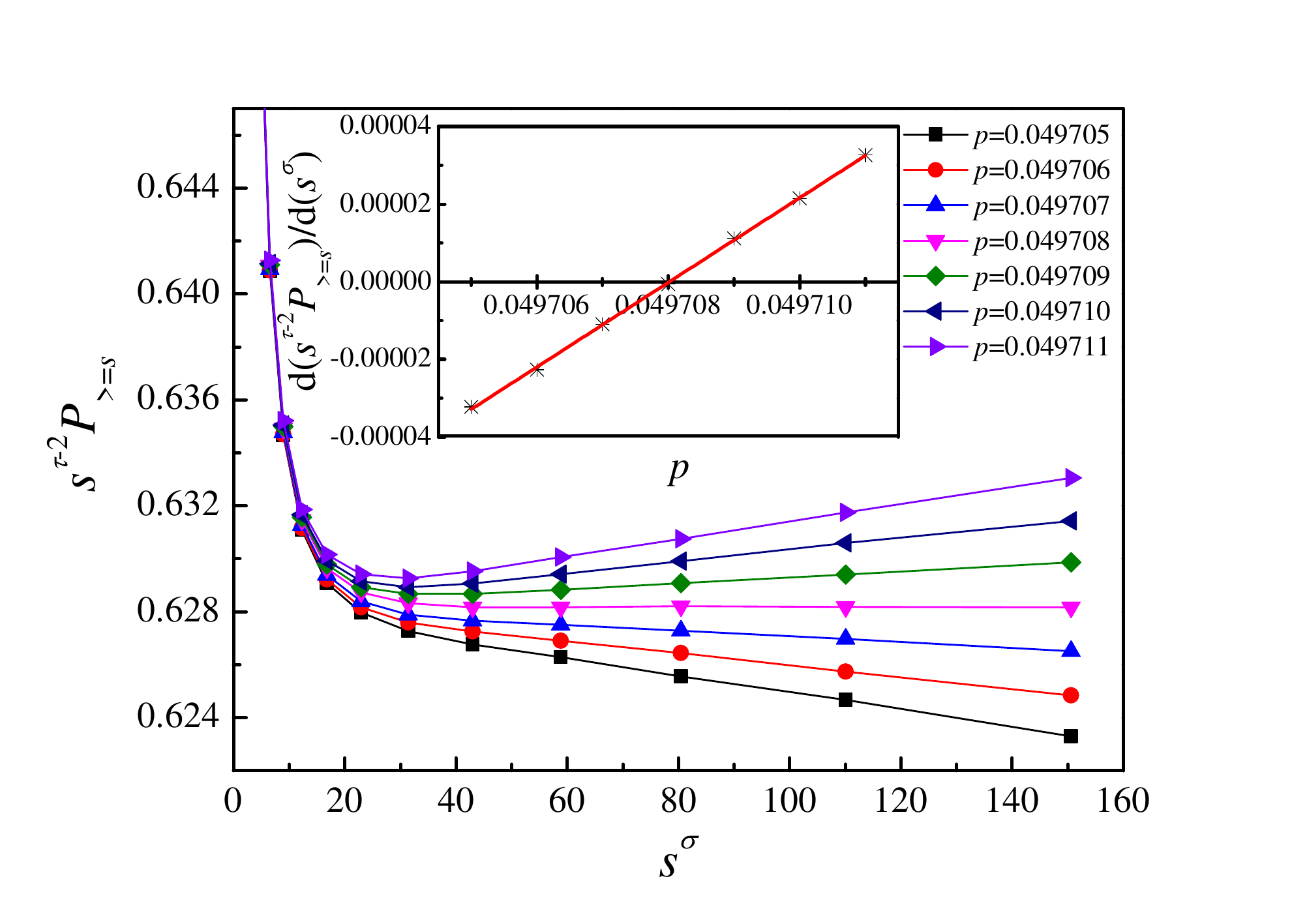} 
\caption{ Plot of $s^{\tau-2}P_{\geq s}$ vs $s^{\sigma}$ with $\tau = 2.18905$ and $\sigma = 0.4522$ for the sc-NN+2NN+3NN lattice under different values of $p$. The inset indicates the slope of the linear portions of the curves shown in the main figure as a function of $p$, and the center value of $p_{c} = 0.0497080$ can be calculated from the $x$ intercept.}
\label{fig:sc-nn+2nn+3nn-sigma}
\end{figure}

\begin{figure}[htbp] 
\centering
\includegraphics[width=3.8in]{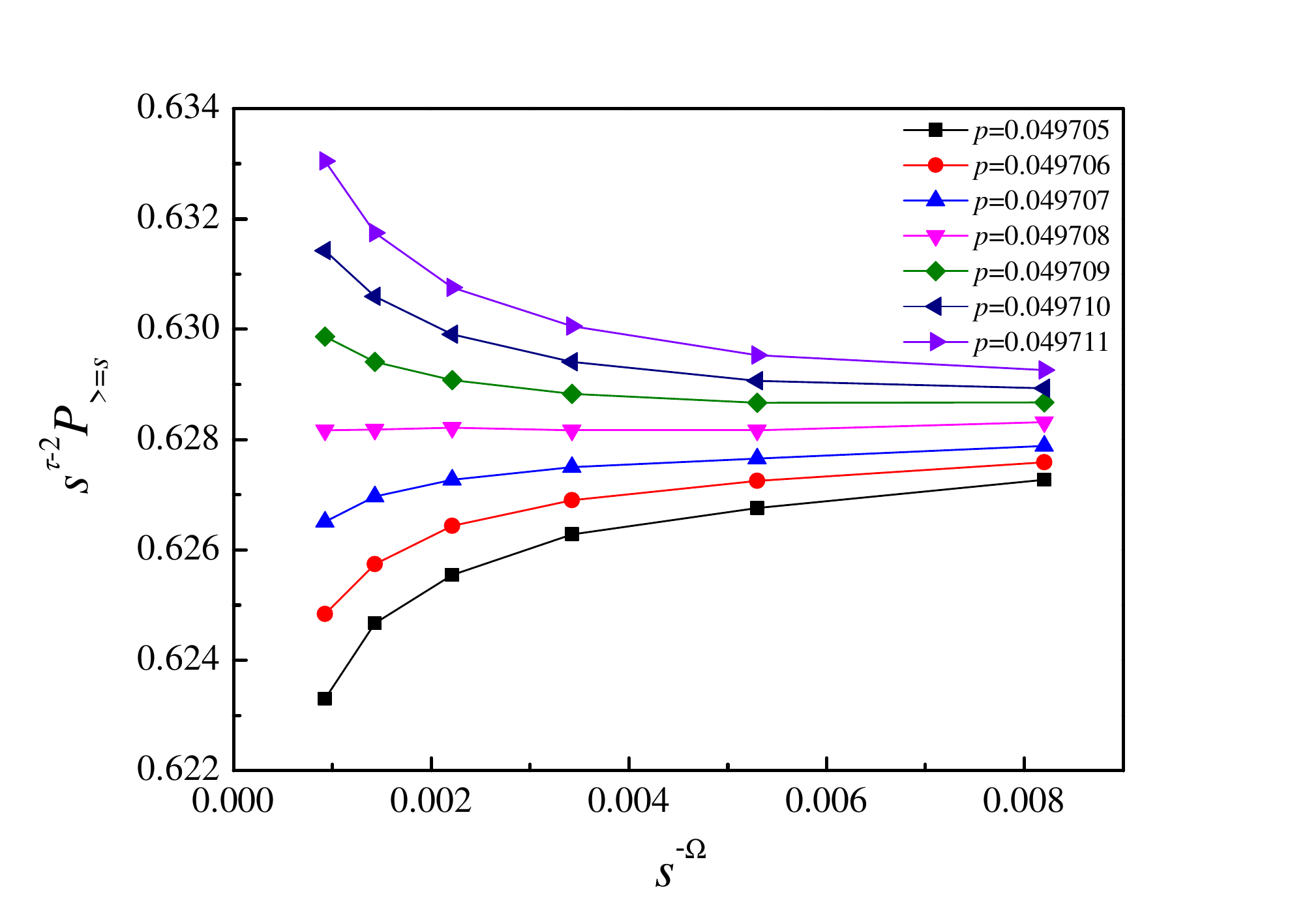} 
\caption{ Plot of $s^{\tau-2}P_{\geq s}$ vs $s^{-\Omega}$ with $\tau = 2.18905$ and $\Omega = 0.63$ for the sc-NN+2NN+3NN lattice under different values of $p$.}
\label{fig:sc-nn+2nn+3nn-omega}
\end{figure}

\begin{figure}[htbp] 
\centering
\includegraphics[width=3.8in]{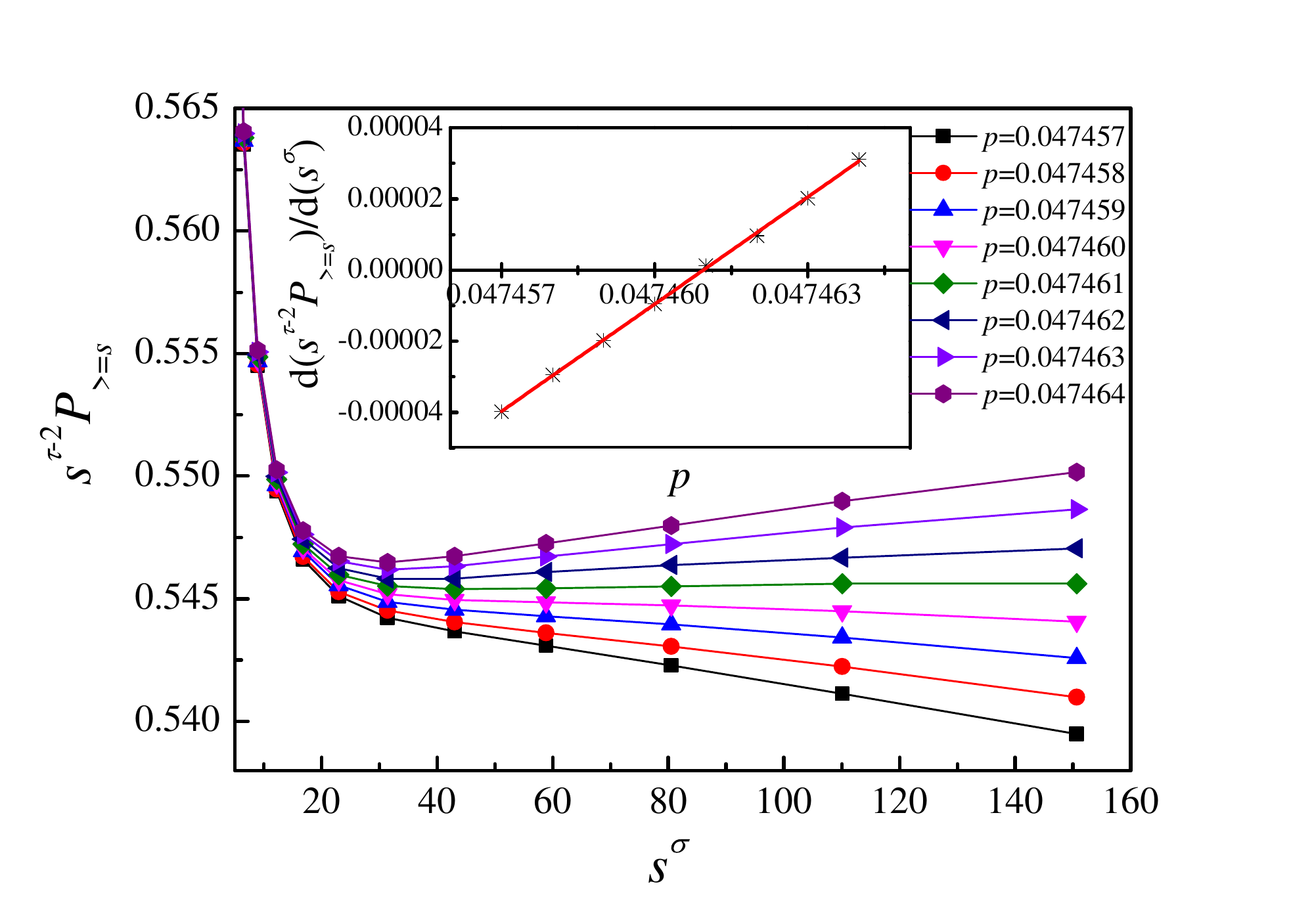} 
\caption{ Plot of $s^{\tau-2}P_{\geq s}$ vs $s^{\sigma}$ with $\tau = 2.18905$ and $\sigma = 0.4522$ for the sc-2NN+3NN+4NN lattice under different values of $p$. The inset indicates the slope of the linear portions of the curves shown in the main figure as a function of $p$, and the center value of $p_{c} = 0.0474609$ can be calculated from the $x$ intercept.}
\label{fig:sc-2nn+3nn+4nn-sigma}
\end{figure}

\begin{figure}[htbp] 
\centering
\includegraphics[width=3.8in]{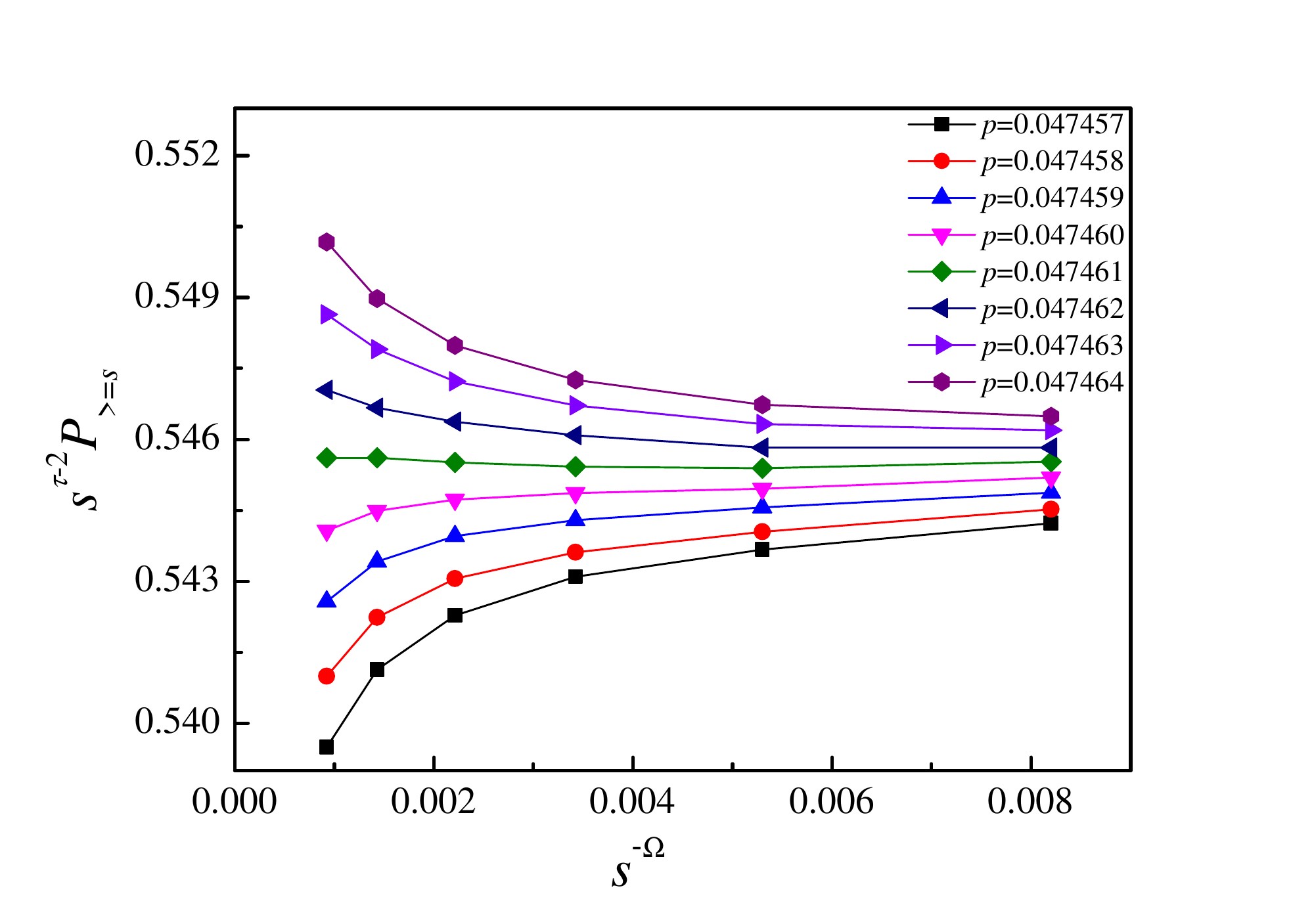} 
\caption{ Plot of $s^{\tau-2}P_{\geq s}$ vs $s^{-\Omega}$ with $\tau = 2.18905$ and $\Omega = 0.63$ for the sc-2NN+3NN+4NN lattice under different values of $p$.}
\label{fig:sc-2nn+3nn+4nn-omega}
\end{figure}

\begin{figure}[htbp] 
\centering
\includegraphics[width=3.8in]{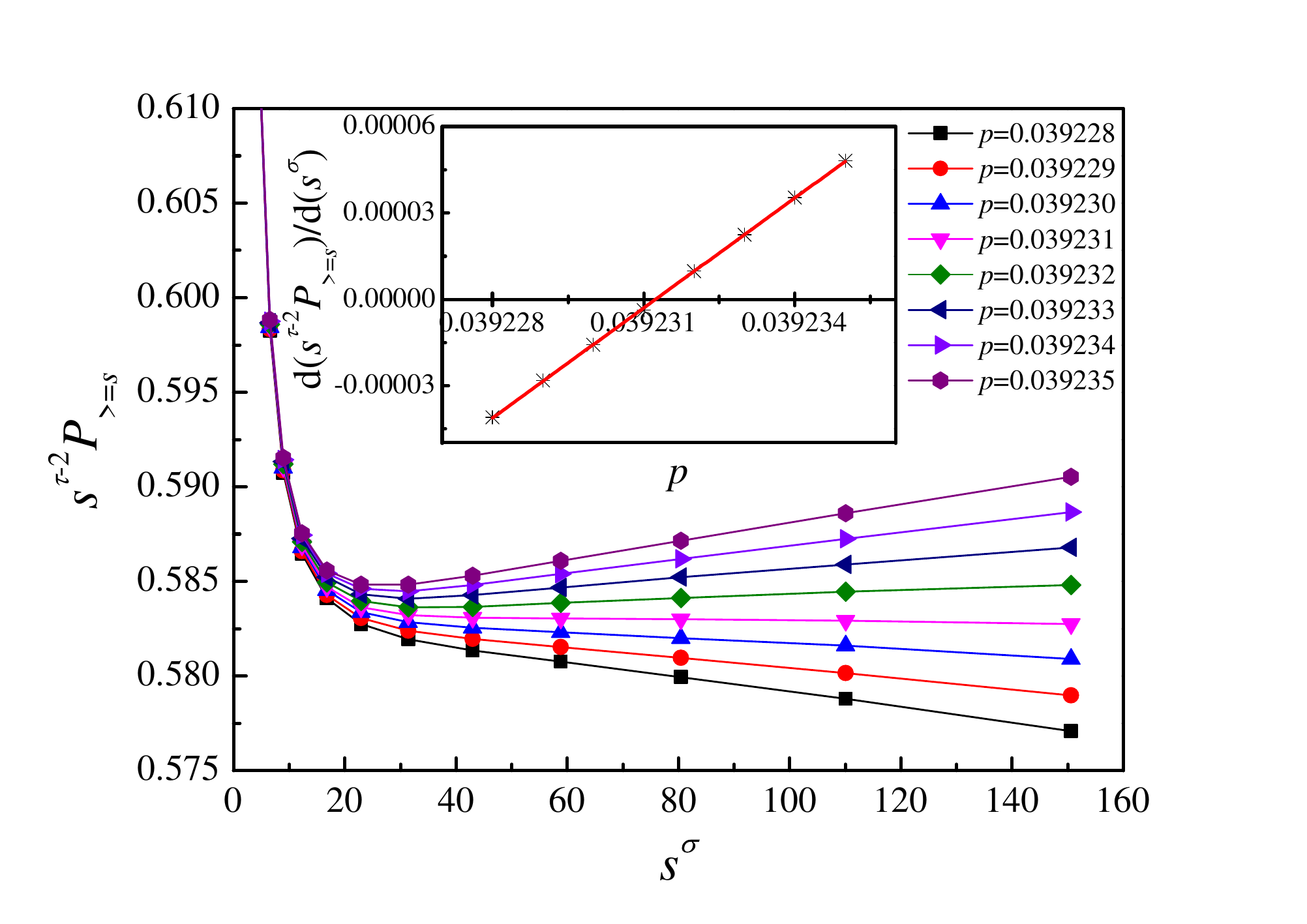} 
\caption{ Plot of $s^{\tau-2}P_{\geq s}$ vs $s^{\sigma}$ with $\tau = 2.18905$ and $\sigma = 0.4522$ for the sc-NN+2NN+3NN+4NN lattice under different values of $p$. The inset indicates the slope of the linear portions of the curves shown in the main figure as a function of $p$, and the center value of $p_{c} = 0.0392312$ can be calculated from the $x$ intercept.}
\label{fig:sc-nn+2nn+3nn+4nn-sigma}
\end{figure}

\begin{figure}[htbp] 
\centering
\includegraphics[width=3.8in]{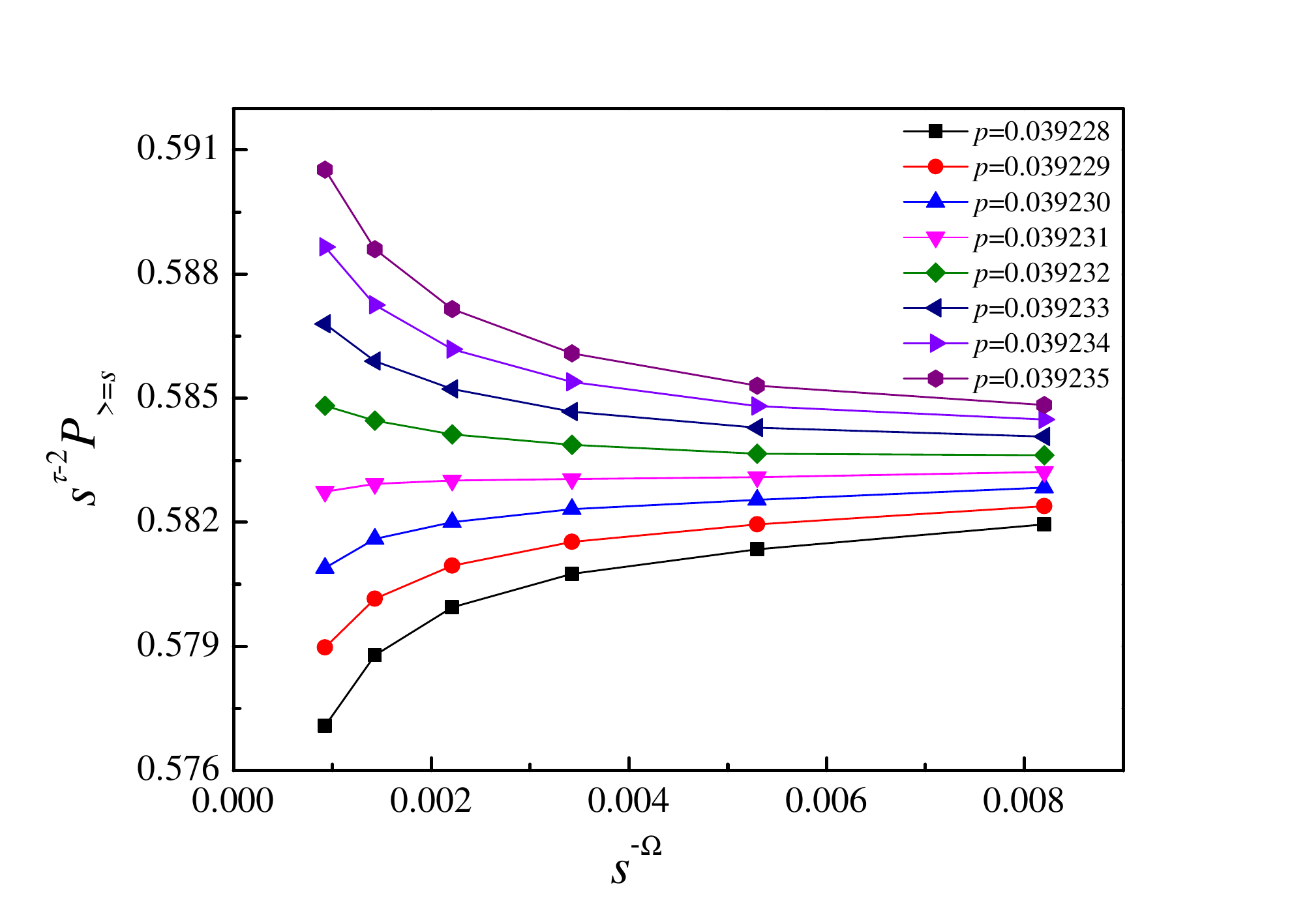} 
\caption{ Plot of $s^{\tau-2}P_{\geq s}$ vs $s^{-\Omega}$ with $\tau = 2.18905$ and $\Omega = 0.63$ for the sc-NN+2NN+3NN+4NN lattice under different values of $p$.}
\label{fig:sc-nn+2nn+3nn+4nn-omega}
\end{figure}

Here we show the results of plots of $s^{\tau-2} P_{\ge s}$  vs $s^{\sigma}$ or $s^{-\Omega}$
for the ten additional lattices not discussed in the main text:
\begin{itemize}
\item sc-3NN+4NN, Figs.\ \ref{fig:sc-3nn+4nn-sigma} and \ref{fig:sc-3nn+4nn-omega}, 
\item sc-NN+3NN, Figs.\ \ref{fig:sc-nn+3nn-sigma} and \ref{fig:sc-nn+3nn-omega},
\item sc-NN+2NN, Figs.\ \ref{fig:sc-nn+2nn-sigma} and \ref{fig:sc-nn+2nn-omega},
\item sc-2NN+4NN, Figs.\ \ref{fig:sc-2nn+4nn-sigma} and \ref{fig:sc-2nn+4nn-omega},
\item sc-2NN+3NN, Figs.\ \ref{fig:sc-2nn+3nn-sigma} and \ref{fig:sc-2nn+3nn-omega},
\item sc-NN+3NN+4NN, Figs.\ \ref{fig:sc-nn+3nn+4nn-sigma} and \ref{fig:sc-nn+3nn+4nn-omega},
\item sc-NN+2NN+4NN, Figs.\ \ref{fig:sc-nn+2nn+4nn-sigma} and \ref{fig:sc-nn+2nn+4nn-omega},
\item sc-NN+2NN+3NN, Figs.\ \ref{fig:sc-nn+2nn+3nn-sigma} and \ref{fig:sc-nn+2nn+3nn-omega},
\item sc-2NN+3NN+4NN, Figs.\ \ref{fig:sc-2nn+3nn+4nn-sigma} and \ref{fig:sc-2nn+3nn+4nn-omega},
\item sc-NN+2NN+3NN+4NN, Figs.\ \ref{fig:sc-nn+2nn+3nn+4nn-sigma} and \ref{fig:sc-nn+2nn+3nn+4nn-omega},
\end{itemize}
The plots for sc-NN+4NN lattice are discussed in the main text in Figs.\ 3 and 4.  The resulting thresholds  are summarized in Table I of the text.
 We did not calculate the apparent values of $\tau$ for all these lattices one by one; otherwise, the overall simulation time would have at least doubled.  For all these plots we assumed the values $\tau = 2.18905$ and $\sigma = 0.4522$.

\clearpage

\end{document}